\documentclass[prd,aps,twocolumn,floatfix,superscriptaddress,nofootinbib]{revtex4-1}

\usepackage[utf8]{inputenc}
\usepackage{amsmath, amsthm, amsfonts, amssymb, latexsym}
\usepackage{graphicx}
\usepackage{color}
\usepackage{subfigure}
\usepackage{url}
\usepackage{psfrag}
\usepackage{mathrsfs}
\usepackage{multirow}
\usepackage{comment}
\usepackage[normalem]{ulem}
\usepackage{tikz} 
\usepackage{bbold}
\usepackage{bbm}
\usepackage{flushend}
\usepackage{mathtools}
\usepackage{slashed}

\usepackage{enumitem}
\setdescription{font=\normalfont}

\def\p{\partial}
\def\Lie{\mathcal{L}}
\DeclareMathAlphabet\mathbfcal{OMS}{cmsy}{b}{n}

\usepackage[unicode]{hyperref}
\hypersetup{
    unicode=true,          
    pdftitle={GaugeBondiHyp},
    pdfcreator={LaTeX},          
    pdfproducer={LaTeX},         
    pdfkeywords={},
    colorlinks=false,            
  }

\DeclareSymbolFont{matha}{OML}{txmi}{m}{it}
\DeclareMathSymbol{\varv}{\mathord}{matha}{118}

\graphicspath{ {./images/} }

\begin{document}

\title{Gauge structure of the Einstein field equations in Bondi-like
  coordinates}

\author{Thanasis Giannakopoulos}
\affiliation{
  Centro de Astrof\'{\i}sica e Gravita\c c\~ao -- CENTRA,
  Departamento de F\'{\i}sica, Instituto Superior T\'ecnico -- IST,
  Universidade de Lisboa -- UL, Av.\ Rovisco Pais 1, 1049-001 Lisboa,
  Portugal}

\author{Nigel T. Bishop}
\affiliation{
  Department of Mathematics, Rhodes University,
  Grahamstown 6140, South Africa
}

\author{David Hilditch}
\affiliation{
  Centro de Astrof\'{\i}sica e Gravita\c c\~ao -- CENTRA,
  Departamento de F\'{\i}sica, Instituto Superior T\'ecnico -- IST,
  Universidade de Lisboa -- UL, Av.\ Rovisco Pais 1, 1049-001 Lisboa,
  Portugal}

\author{Denis Pollney}
\affiliation{
  Department of Mathematics, Rhodes University,
  Grahamstown 6140, South Africa
}

\author{Miguel Zilh\~ao}
\affiliation{
  Centro de Astrof\'{\i}sica e Gravita\c c\~ao -- CENTRA,
  Departamento de F\'{\i}sica, Instituto Superior T\'ecnico -- IST,
  Universidade de Lisboa -- UL, Av.\ Rovisco Pais 1, 1049-001 Lisboa,
  Portugal}
\affiliation{
  Departamento de Matem\'atica da Universidade de Aveiro and Centre
  for Research and Development in Mathematics and Applications
  (CIDMA), Campus de Santiago, 3810-183 Aveiro, Portugal
}

\begin{abstract}
  The characteristic initial (boundary) value problem has numerous
  applications in general relativity (GR) involving numerical studies
  and is often formulated using Bondi-like coordinates. Recently it
  was shown that several prototype formulations of this type are only
  weakly hyperbolic. Presently we examine the root cause of this
  result. In a linear analysis we identify the gauge, constraint and
  physical blocks in the principal part of the Einstein field
  equations (EFE) in such a gauge, and we show that the subsystem
  related to the gauge variables is only weakly hyperbolic. Weak
  hyperbolicity of the full system follows as a consequence in many
  cases. We demonstrate this explicitly in specific examples, and thus
  argue that Bondi-like gauges result in weakly hyperbolic free
  evolution systems under quite general conditions. Consequently the
  characteristic initial (boundary) value problem of GR in these
  gauges is rendered ill-posed in the simplest norms one would like to
  employ. The possibility of finding good alternative norms, in which
  well-posedness is achieved, is discussed. So motivated, we present
  numerical convergence tests with an implementation of full GR which
  demonstrate the effect of weak hyperbolicity in practice.
\end{abstract}

\maketitle

\section{Introduction} \label{Section:Intro}

Characteristic formulations of general relativity (GR) have proven to
be particularly useful in a number of cases. In the growing field of
gravitational wave astronomy, they can help provide waveform models
with high accuracy. Since characteristic formulations are based on
null hypersurfaces, future null infinity can be naturally included in
the computational domain. This is the region where quantities such as
the Bondi news function are unambiguously defined, and so methods such
as Cauchy characteristic extraction (CCE)~\cite{BisGomLeh96a,
  BisGomLeh97a, ZloGomHus03, BabSziHaw05, GomBarFri07, BabBisSzi09,
  ReiBisPol09, ReiBisPol09a, BabSziWin11, BabWinZlo11, HanSzi14,
  BarMoxSch19, MoxSchTeu20, Ioz21, Mit21, IozBoyDep21, MitMoxSch21,
  Fou21, MoxSchTeu21} and matching (CCM)~\cite{Win12, Szi00} can
eliminate systematic extrapolation errors (the main alternative
strategy for this is to use compactified hyperboloidal
slices~\cite{MonRin08,BarSarBuc11,Zen10,VanHusHil14,DouFra16,HilHarBug16,
  VanHus17, SteFri82}).

Characteristic formulations are used more broadly. For instance
characteristic codes have been built to explore the behavior of
relativistic stars~\cite{PapFon99, SieFonMul02}. In the study of
gravitational collapse, codes based on null foliations offer a
practical alternative to the standard spacelike foliation
approach. Their advantage lies in the compactness of the system of
partial differential equations (PDE) solved~\cite{Gar95, CreOliWin19,
  GunBauHil19, SieFonMue03, AlcBarOli21} as well as the inclusion of
null infinity in the computational domain. The aforementioned setups
are usually considered in asymptotically flat geometries; though
see~\cite{OliSop16, Oli17b} for gravitational collapse in
asymptotically anti-de Sitter (AdS) spacetimes. In these geometries,
characteristic formulations have most often been used in the field of
numerical holography. Exploiting holographic duality~\cite{Mal97,
  Wit98}, the aim is to obtain a better insight of the behavior of
strongly coupled matter out of equilibrium~\cite{CheYaf11, CheYaf15,
  JanJanSol17, AttCasMat17a, AttCasMat17b, AttBeaCas18, AttBeaCas19,
  BelJanJan19, WaeRabSch19, BeaDiaGia21, BeaCasGia21, BeaCasGia21a}
(for an introduction see the reviews~\cite{CheYaf14, LiuSon18}). Even
applications to cosmology have been pursued~\cite{vdWBis12}.

When formulating the characteristic initial value problem (CIVP) or
characteristic initial boundary value problem (CIBVP) in GR, it is
common practice to choose a Bondi-like gauge. Codes built upon these
formulations have successfully passed a plethora of tests and provided
physically sensible results. Their performance and stability during
simulations has often led to the expectation that the continuum PDE
problem is well-posed. A PDE problem is called well-posed if it
possesses a unique solution that depends continuously, in an
appropriate norm, on the given data. The existence and uniqueness of
solutions to the Bondi-like CIBVP in GR have long been
studied~\cite{FriLeh99, GomFri03}. Recently, working with first order
reductions, continuous dependence on given data was examined by
analyzing the degree of hyperbolicity of the continuum PDE
systems~\cite{GiaHilZil20}. These reductions can be written in the
compact form
\begin{align*}
  \mathbfcal{A}^t(\mathbf{u},x^\mu) \, \p_t \mathbf{u}
  + \mathbfcal{A}^p(\mathbf{u},x^\mu)\,
  \p_p \mathbf{u} +  \mathbfcal{S}(\mathbf{u},x^\mu) = 0
  \,,
\end{align*}
with the state vector~$\mathbf{u} = (u_1, u_2, \dots, u_q )^T$ and
principal part matrices,
\begin{align*}
  \mathbfcal{A}^\mu =
  \begin{pmatrix}
    a^{\mu}_{11}  & \dots  & a^{\mu}_{1q} \\
    \vdots & \ddots & \vdots \\
    a^{\mu}_{q1}  & \dots  & a^{\mu}_{qq} 
  \end{pmatrix}
\end{align*}
where~$\det(\mathbfcal{A}^t) \neq 0 \, $. Working in the constant
coefficient approximation, the degree of hyperbolicity of the system
can be classified locally by examining the principal symbol
\begin{align*}
  \mathbf{P}^s = \left(\mathbfcal{A}^t \right)^{-1}
  \mathbfcal{A}^p \, s_p \,,
\end{align*}
with~$s^i$ an arbitrary unit spatial vector. The PDE system is called
\textit{weakly hyperbolic}~(WH) if the principal symbol has real
eigenvalues for all~$s^i$. It is called \textit{strongly
  hyperbolic}~(SH) if moreover~$\mathbf{P}^s$ is diagonalizable for
all~$s^i$ and a constant~$K$ independent of~$s^i$ that satisfies
\begin{align*}
|\mathbf{T}_s|+|\mathbf{T}_s^{-1}|\leq K,
\end{align*}
exists, with~$\mathbf{T}_s$ the similarity matrix that
diagonalizes~$\mathbf{P}^s$.

A classic strategy~\cite{Ren90,Fri05} for well-posedness analysis of
the CIVP is to reduce to an initial value problem
(IVP). Well-posedness in~$L^2$ for the IVP is characterized by strong
hyperbolicity~\cite{GusKreOli95,Hil13}. The IVP of a WH PDE system is
ill-posed in~$L^2$, but it may be well-posed in a different
norm~\cite{KreLor89}. This well-posedness is, however, delicate and,
unlike the well-posedness of a SH PDE, can be broken by source
terms. Well-posedness is a necessary condition for a numerical
approximation of a PDE problem to converge to the continuum solution
with increasing resolution. Convergence here is to be understood in
terms of a discretized version of the norm in which the continuum
problem is well-posed. An error estimate obtained from the numerical
solutions of an ill-posed PDE problem should,~\textit{a priori}, be
treated with great care. Therefore, well-posedness of the Bondi-like
CIVP and CIBVP in GR is a particularly pressing open question for
studies that focus on accuracy.

The result of~\cite{GiaHilZil20} was that two commonly used Bondi-like
gauges give rise to second order PDE systems that are only WH outside
of the spherical context. Toy models that mimic this structure were
used to demonstrate the effect of weak hyperbolicity in numerical
experiments. In this paper we examine the cause of this result and,
following~\cite{KhoNov02, HilRic13}, identify this weak hyperbolicity
as a pure gauge effect. We argue that the construction upon radial
null geodesics renders the vacuum Einstein field equations (EFE) only
WH in all Bondi-like gauges. We explicitly show the effect of weak
hyperbolicity in numerical experiments in full GR formulated in the
Bondi-Sachs proper gauge.  This result implies that the CIVP and CIBVP
of GR in vacuum are ill-posed in the simplest norms one might like to
employ when formulated in these gauges. This issue can potentially be
sidestepped by working with alternative norms, or higher derivatives
of the metric, which might be taken explicitly as evolved variables,
or simply placed within the norm under consideration. The latter tack
has been successfully followed in, for
example,~\cite{Rac13,CabChrTag14,HilValZha19,Rip21}.

In Sec.~\ref{Section:BS_to_ADM} we map the Bondi-like equations and
variables to the Arnowitt-Deser-Misner (ADM) ones, so that we can
straightforwardly apply the aforementioned tools. In
Sec.~\ref{Section:princ_symbol} we summarize the relevant theory and
the structure of the principal part resulting from gauge freedom. In
Sec.~\ref{Section:aff_null_gauge} we analyze the affine null
gauge~\cite{Win12}, showing it to be only WH, both in the
characteristic and in the equivalent ADM setups. In
Sec.~\ref{Subsection:BS_proper_gauge} this analysis is repeated for
the Bondi-Sachs gauge proper~\cite{BonBurMet62,Sac62} in the ADM
setup. In Sec.~\ref{Subsection:double_null_gauge} the calculation is
repeated for the double null gauge~\cite{Chr08}. We argue that all
Bondi-like gauges possess this pure gauge structure.  In
Sec.~\ref{Section:numerics} we examine the numerical conseqeuences of
WH by performing robust-stability-like~\cite{BabHusAli08, AlcAllBon03,
  CalHinHus05} tests. The results are compared against those of an
artificial SH system. The tests are performed using the publicly
available~\texttt{PITTNull} thorn of the~\texttt{Einstein
  Toolkit}~\cite{einsteintoolkitzenodo}. We conclude in
Sec.~\ref{Section:Conclusions}. Geometric units are used
throughout. Our scripts are available in the ancillary files and our
data in~\cite{GiaBisHil21_public}.

\section{Bondi-like formulations and their ADM
  equivalent}\label{Section:BS_to_ADM}

In this section we review the main features of Bondi-like formulations
and map the corresponding equations and variables to the ADM
language. To do so we employ a coordinate transformation between
generalized Bondi-like and ADM coordinates. The gauge and thus the PDE
character of the system are fixed by the Bondi-like coordinates. This
choice determines, for instance, which metric components and/or
derivatives thereof vanish. The subsequent transformation to the ADM
coordinates merely results in relabeling variables and expressing
directional derivatives of the Bondi-like basis in terms of those of
the ADM basis.

\subsection{Main features of Bondi-like formulations}
\label{Subsection:BS-like_features}

To demonstrate relevant features common to all Bondi-like gauges we
work with the generalized Bondi-Sachs formulation of~\cite{CaoHe13}
with line element
\begin{equation}
  \begin{aligned}
    ds^2
    & = g_{uu} du^2 + 2 g_{ur} du\,dr
    + 2 g_{u \theta} du \, d\theta
    + 2 g_{u \phi} du \, d\phi
    \\
    & \qquad \qquad \quad \; \; \,
    + g_{\theta \theta} d\theta^2
    + 2 g_{\theta \phi} d\theta \, d\phi
    + g_{\phi \phi} d \phi^2
    \,.
  \end{aligned}
  \label{eqn:gen_BS_line_element}
\end{equation}
We consider a four-dimensional spacetime and identify the
coordinates~$\theta,\phi$ with the usual spherical polar angles on the
two-sphere. All seven nontrivial metric components
of~\eqref{eqn:gen_BS_line_element} are functions of the characteristic
coordinates~$x^{\mu'}=(u,r,\theta,\phi)$, with the hypersurfaces of
constant~$u$ null and henceforth denoted by~$\mathcal{N}_u$. The null
vector~$(\p/\p_r)^a$ is both tangent and normal to~$\mathcal{N}_u$ and
hence orthogonal to the spatial vectors~$(\p/\p_\theta)^a$
and~$(\p/\p_\phi)^a$ that lie within~$\mathcal{N}_u$. This vector
basis guarantees that
\begin{align}
  g^{uu} = g^{u\theta} = g^{u\phi} = 0
  \,,
  \label{eqn:gauge_cond_null}
\end{align}
and every distinct null geodesic in~$\mathcal{N}_u$ can be labeled
by~$\theta,\phi$. The characteristic hypersurface~$\mathcal{N}_u$ can
be either outgoing or ingoing. If the formulation incorporates both
types of null hypersurfaces, then the double null gauge~\cite{Chr08}
is imposed. In this case~$g^{rr}=0$ and the coordinates~$u,r$
correspond to the advanced and retarded time rather than an advanced
(or retarded) time and the radial coordinate.

A free evolution PDE system for the vacuum EFE in an asymptotically
flat spacetime in a Bondi-like gauge consists of
\begin{align}
  R_{rr} = R_{r\theta} = R_{r\phi} = R_{\theta \theta} = R_{\theta \phi}
  = R_{\phi \phi} = 0
  \,,
  \label{eqn:main_BS_sys}
\end{align}
which is often called the main system. The equation~$R_{ur}=0$ is
commonly referred to as the trivial equation, because solutions to the
main system automatically satisfy it, as shown in~\cite{BonBurMet62,
  CaoHe13} via the contracted Bianchi identities. The supplementary
equations
\begin{align*}
  R_{uu} = R_{u\theta} = R_{u\phi} = 0
\end{align*}
are guaranteed to be satisfied in~$\mathcal{N}_{u}$ if they are
satisfied on a cross section~\cite{BonBurMet62, CaoHe13}. The main
system provides six evolution equations for the seven unknown metric
functions. Usually, a definition for the determinant of the induced
metric on the two-spheres is made, namely
\begin{align}
  g_{\theta \theta} \, g_{\phi \phi} - g_{\theta \phi}^2
  = \hat{R}^4 \sin^2\theta
  \,,
  \label{eqn:gauge_cond_det}
\end{align}
where~$\hat{R}$ is taken to be a function of the coordinates and
reduces to the areal radius of the two-sphere in spherical symmetry.

The aforementioned are common to all Bondi-like gauges. There is a
residual gauge freedom which corresponds to the choice of the
coordinate labeling the position within the null geodesic. This is
done differently in the various Bondi-like gauges. We focus on three
common choices:
\begin{description}  
\item[\textit{Affine null}~\cite{Win13, CreOliWin19}] The final choice
  of equations is achieved by setting~$g^{ur}= - 1$ for
  outgoing~$\mathcal{N}_u$ and~$g^{ur}= 1$ for
  ingoing~$\mathcal{N}_u$.~$\hat{R}$ is then taken to be an unknown of
  the problem.
\item[\textit{Bondi-Sachs proper}~\cite{BonBurMet62}] The radial
  coordinate matches the areal radius~$\hat{R}=r$ and so the
  definition~\eqref{eqn:gauge_cond_det} reduces the number of unknowns
  to six.
\item[\textit{Double null}~\cite{Chr08}] The residual gauge freedom is
  fixed by the condition~$g^{rr}=0$.
\end{description}

\subsection{From the characteristic to the ADM equations}
\label{Subsection:BS_to_ADM_eqs}

We now map from the characteristic to the ADM variables and present
the system equivalent to~\eqref{eqn:main_BS_sys} in the ADM
formalism. We assume that~$\mathcal{N}_u$ are outgoing, but an
analogous analysis can be performed for ingoing null hypersurfaces. To
begin, we choose ADM coordinates~$x^{\mu}=(t,\rho,\theta,\phi)$. They
are related to the characteristic coordinates via
\begin{align}
  u = t - f(\rho) \,, \quad r = \rho
  \,.
  \label{eqn:coord_transf}
\end{align}
As in~\cite{Fri06}, the quantity~$-df/dr$ determines the slope of the
constant~$t$ spacelike hypersurface~$\Sigma_t$ on the~$u,r$ plane. The
angular coordinates~$\theta,\phi$ are unchanged, and in this
subsection we may label them with the Latin indices~$A,B$.

The lapse of proper time between~$\Sigma_t$ and~$\Sigma_{t+dt}$ along
their normal observers is~$d \tau = \alpha(t,x^i) dt$, with the lapse
function defined by
\begin{align*}
  \alpha^{-2}(t,x^i)  \equiv -
  g^{\mu \nu} \nabla_\mu t
  \nabla_\nu t
  \,.
\end{align*}
The relative velocity between the trajectory of those observers and
the lines of constant spatial coordinates is given
by~$\beta^i(t,x^j)$, where~$x^i_{t+dt} = x^i_t -
\beta^i(t,x^j)dt$. The quantity~$\beta^i$ is called the shift vector.
The future directed unit normal 4-vector on~$\Sigma_t$ is
\begin{align*}
  n^{\mu} \equiv - \alpha \nabla^{\mu} t
  =
  \alpha^{-1} \left(1, -\beta^i \right)
  \,,
\end{align*}
and its covector form is
\begin{align*}
  n_{\mu} = g_{\mu \nu} n^{\nu}
  =
   \left(-\alpha, 0,0,0 \right)
  \,.
\end{align*}
The metric induced on~$\Sigma_t$ is
\begin{align*}
  \gamma_{\mu\nu} \equiv g_{\mu\nu} + n_\mu n_\nu
  \,.
\end{align*}
The ADM form of the equations is obtained by systematic contraction
with~$n^{\mu}$ and~$\gamma_{\mu \nu}$. This geometric construction is
discussed in most numerical relativity textbooks~\cite{Alc08, Gou07,
  BauSha10}. The spacetime metric takes the form
\begin{align*}
  g_{\mu \nu} =
  \begin{pmatrix}
    -\alpha^{2} + \beta_k \beta^k & \beta_{i} \\
    \beta_{j}  & \gamma_{i j} \\
  \end{pmatrix}\,,
\end{align*}
where lowercase Latin indices denote spatial components. The inverse
of~$g_{\mu \nu}$ is
\begin{align*}
  g^{\mu \nu} =
  \begin{pmatrix}
    -\alpha^{-2}  & \alpha^{-2} \beta^{i} \\
    \alpha^{-2} \beta^{j}  & \gamma^{i j}
    - \alpha^{-2} \beta^{i} \beta^{j} \\
  \end{pmatrix}\,.
\end{align*}
By comparing the~$3+1$ form of the metric and its inverse to the
generalized Bondi version~\eqref{eqn:gen_BS_line_element} we can
interpret the Bondi-like gauges in terms of lapse and shift, and
relate the characteristic variables to the ADM ones. Every Bondi-like
vector basis gives~\eqref{eqn:gauge_cond_null}, which in ADM
coordinates reads
\begin{equation*}
  \begin{aligned}
    g^{uu}
    &= \frac{\p u}{\p x^\mu}  \frac{\p u}{\p x^\nu} g^{\mu \nu}
    = g^{tt} -2 f'g^{t \rho} + (f')^2 g^{\rho \rho} = 0
    \,,
    \\
    g^{uA}
    &= \frac{\p u}{\p x^\mu}  \frac{\p x^A}{\p x^\nu} g^{\mu \nu}
    = g^{tA} - f'g^{\rho A} = 0
    \,,
  \end{aligned}
\end{equation*}
and leads to
\begin{equation}
  \begin{aligned}
    \gamma^{\rho \rho}
    &= \left(\frac{1 + f' \beta^\rho}{f' \alpha}\right)^2
    \,,
    \quad
    \gamma^{\rho A}
     = \beta^A \frac{1 + f' \beta^\rho }{f' \alpha^2}
    \,.
  \end{aligned}
  \label{eqn:gauge_cond_null_gamma_uu}
\end{equation}
The Bondi-like metric ansatz \eqref{eqn:gen_BS_line_element} implies
\begin{align*}
  g_{rr} = g_{r A} = 0
  \,,
\end{align*}
which after using~$\beta_i = \gamma_{ij} \beta^j$ yields
\begin{equation}
  \begin{aligned}
    \gamma_{\rho \rho}
    &= \frac{(f')^2 (\alpha^2 + \beta^A \beta^B \gamma_{AB})}
    {(1 + f' \beta^\rho)^2}
    \,,
    \\
    \gamma_{\rho A}
    &= - \frac{f' }{1 + f' \beta^\rho} \beta^B \gamma_{AB} 
    \,.
  \end{aligned}
  \label{eqn:gauge_cond_null_gamma_dd}
\end{equation}
Using the latter
and~$g_{\mu^\prime \nu^\prime} = \frac{\p x^\mu}{\p x^{\mu^\prime}}
\frac{\p x^\nu}{\p x^{\nu^\prime}} g_{\mu \nu}$ provides the following
relations between the characteristic and ADM variables, for all
Bondi-like gauges:
\begin{equation}
 \begin{aligned}
   g_{uu}
   & = \frac{\beta^A \beta_A -\alpha^2 (1 + 2 f' \, \beta^\rho)}
   {(1 + f' \, \beta^\rho)^2}
   \,,&
   g_{ur} &
   = \frac{ - f' \, \alpha^2}{1+ f' \beta^\rho}
   \,,
   \\
   g_{uA} & = - \gamma_{\rho A}/f' \,, &
   g_{AB} & = \gamma_{AB} \,.
 \end{aligned}
 \label{eqn:gen_BS_vars_to_ADM}
\end{equation}
The above combined
with~$\gamma^{AB} - \alpha^{-2} \beta^A \beta^B = g^{AB}$ further
yield
\begin{align}
  &\gamma^{\theta \theta}
    = \left( \frac{\beta^\theta}{\alpha}\right)^2
    + \frac{\gamma_{\phi \phi}}{\det(g_{AB})}
    \,, \quad
    \gamma^{\theta \phi}
    = \frac{\beta^\theta \beta^\phi}{\alpha^2}
    - \frac{\gamma_{\theta \phi}}{\det(g_{AB})}
    \,,
    \nonumber
  \\
  &
    \gamma^{\phi \phi}
    = \left( \frac{\beta^\phi}{\alpha}\right)^2
    + \frac{\gamma_{\theta \theta}}{\det(g_{AB})}
    \label{eqn:gauge_cond_null_gamma_uu_sphere}
\end{align}
for all Bondi-like gauges.

To proceed with the mapping between characteristic and ADM formalism,
we simply take the standard tensor transformation rule. The main
system~\eqref{eqn:main_BS_sys} written in the ADM coordinates is then
\begin{equation}
  \begin{aligned}
    R_{rr} & = (f')^2 R_{tt} + 2 f' R_{t \rho} + R_{\rho \rho} = 0 
    \,,
    \\
    R_{r A} &= f' R_{t A} + R_{\rho A} = 0 
    \,,
    \\
    R_{AB} & = 0
    \,.
  \end{aligned}
  \label{eqn:main_BS_in_ADM_coords}
\end{equation}
The complete orthogonal projection onto~$\Sigma_t$ is given by
\begin{align}
  \gamma^{\lambda}{}_{\mu} \gamma^{\sigma}{}_{\nu} R_{\lambda \sigma}
  \equiv R^\perp_{\mu \nu}
  &=
    -\mathcal{L}_n K_{\mu \nu}
    - \frac{1}{\alpha} D_{\mu} D_{\nu} \alpha
    + {}^{(3)}R_{\mu \nu}
    \nonumber
  \\
  &
    + K K_{\mu \nu}
    -2 K_{\mu \lambda} K^{\lambda}{}_{\nu}
    \,,
    \label{eqn:gamma_gamma_Rdd}
\end{align}
with~$R^\perp_{\mu \nu}$ a purely spatial tensor, and
\begin{align}
  \gamma^{\mu}{}_{\nu}
  &= \delta^{\mu}{}_{\nu} + n^{\mu} n_{\nu}
  \,,
  \label{eqn:projector_general}
  \\
  K_{\mu \nu}
  & =
  - \left( \nabla_{\mu} n_{\nu}
  + n_{\mu} n^{\kappa} \nabla_{\kappa} n_{\nu} \right)
  \,,
  \nonumber
\end{align}
the orthogonal projector and the extrinsic curvature of~$\Sigma_t$
when embedded in the full spacetime, respectively. The following
purely spatial quantities have been used:
\begin{align*}
   D_{\mu} S_{\nu \lambda}
  &=
    \perp \nabla_\mu  S_{\nu \lambda}  
    \,,
  \, \qquad 
  {}^{(3)}\Gamma^\mu{}_{\nu \lambda}
  =
    \perp \Gamma^\mu{}_{\nu \lambda}
    \,,
  \\
  {}^{(3)}R_{\mu \nu}
  & =
    \perp \left(
    \p_\lambda {}^{(3)}\Gamma^\lambda{}_{\mu \nu}
    -
    \p_\nu {}^{(3)}\Gamma^\lambda{}_{\mu \lambda}
    \right.
  \\
  & \quad \left.
    +
    {}^{(3)}\Gamma^\lambda{}_{\mu \nu}  {}^{(3)}\Gamma^\sigma{}_{\lambda \sigma}
    -
    {}^{(3)}\Gamma^\lambda{}_{\mu \sigma}  {}^{(3)}\Gamma^\sigma{}_{\nu \lambda}
    \right)
    \,,
\end{align*}
where~$D_\mu$ is the covariant derivative compatible
with~$\gamma_{\mu \nu}$, the symbol~$\perp$ denotes projection
with~$\gamma^\mu{}_\nu$ on every open index and~$S_{\mu \nu}$ denotes
an arbitrary spatial tensor. Imposing~$R_{\mu \nu} = 0$ and focusing
only on the spatial components of~$R^\perp_{\mu \nu}$ one can obtain
the evolution equations for the spatial components of the extrinsic
curvature
\begin{align*}
  \mathcal{K}_{ij}
  & \equiv - \p_t K_{ij}
    -D_i D_j \alpha
    + \alpha
    \left(
    {}^{(3)}R_{ij} + K K_{ij} -2 K_{im} K^{m}{}_j 
    \right)
    \nonumber
  \\
  &
    + \beta^m \p_m K_{ij}
    + K_{im} \p_j \beta^m
    + K_{mj} \p_i \beta^m
    = 0
    \,,
\end{align*}
where~$K=g^{\mu \nu} K_{\mu \nu}$. The full projection perpendicular
to~$\Sigma_t$ is
\begin{align*}
  n^\mu n^\nu R_{\mu \nu}
  \equiv R^\parallel
  & =
    \mathcal{L}_n K + \frac{1}{\alpha} D^i D_i \alpha - K_{ij} K^{ij}
    \,.
\end{align*}
Using
\begin{align*}
  \mathcal{L}_n K = \gamma^{ij} \mathcal{L}_n K_{ij} + 2 K_{ij} K^{ij}
  \,,
\end{align*}
Eq.~\eqref{eqn:gamma_gamma_Rdd} and imposing the EFE,~$R^\parallel$
provides the Hamiltonian constraint
\begin{align*}
  H \equiv {}^{(3)}R + K^2 - K_{ij} K^{ij}
  = R^\parallel + \gamma^{ij} R^\perp_{ij} = 0
  \,.
\end{align*}
Finally, the mixed projection is given by the contracted Codazzi
relation
\begin{align*}
  n^\mu \gamma^\lambda{}_\nu R_{\mu \lambda}
  \equiv R^{| \perp}_\nu
  &=
    D_\nu K - D_\mu K^\mu{}_\nu 
    \,,
\end{align*}
with~$n^\mu R^{| \perp}_\mu = 0$. After imposing the EFE it yields the
momentum constraints
\begin{align*}
  M_i \equiv 
  D_j K^j{}_i - D_i K
  = 0
  \,.
\end{align*}
From Eq.~\eqref{eqn:projector_general} and the previous projections we
write
\begin{align}
  \delta^\alpha{}_\mu \delta^\beta{}_\nu R_{\alpha \beta} =
  R_{\mu \nu}^\perp + n_\mu n_\nu R^\parallel
  - n_\mu R_\nu^{| \perp} - n_\nu R_\mu^{| \perp}
  \,.
  \label{eqn:Rmunu_in_ADM_proj}
\end{align}
Using Eq.~\eqref{eqn:Rmunu_in_ADM_proj}, with
Eq.~\eqref{eqn:main_BS_in_ADM_coords} and taking linear combinations
of Eq.~\eqref{eqn:main_BS_sys}, we obtain the ADM system
\begin{align}
  &
    \frac{((f')^2-1)(1+f' \beta^\rho)^2}{(f')^2} \mathcal{K}_{\rho \rho} 
    + \alpha^2 H
    - 2 \alpha f' (1 + f' \beta^\rho) M_\rho
    \nonumber
  \\
  & \qquad \qquad \qquad \qquad \qquad \quad \;
    - 2 \alpha \beta^A M_A
    = 0
    \,,
    \nonumber
  \\
  &
    (1 + f' \beta^\rho) \mathcal{K}_{\rho A} - \alpha f' M_A = 0
    \,,
    \label{eqn:equiv_ADM_to_main_sys}
  \\
  &
    \mathcal{K}_{AB} = 0
    \,, \nonumber
\end{align}
which is equivalent to the main Bondi-like
system~\eqref{eqn:main_BS_sys}, where we have also used
Eqs.~\eqref{eqn:gauge_cond_null_gamma_uu},~\eqref{eqn:gauge_cond_null_gamma_dd},
and~\eqref{eqn:gauge_cond_null_gamma_uu_sphere}.

If the slope of~$\Sigma_t$ of the~$3+1$ foliation in the~$u,r$ plane
is~$f' \neq 1$, then the main Bondi-like
system~\eqref{eqn:main_BS_sys} corresponds to evolution equations for
all the components of~$K_{ij}$ with specific addition of the ADM
Hamiltonian and momentum constraints. For~$f'=1$ though, the first
equation of~\eqref{eqn:equiv_ADM_to_main_sys} involves only ADM
constraints. In this foliation the evolution equation
for~$K_{\rho \rho}$ is provided by the trivial equation, which after
imposing~\eqref{eqn:equiv_ADM_to_main_sys} reads
\begin{align*}
  (1+\beta^\rho) \mathcal{K}_{\rho \rho} - \alpha M_\rho
  + \frac{\alpha }{1 + \beta^\rho}  \beta^A M_A
  =0
  \,.
\end{align*}

The lapse and shift are not determined by the Einstein equations, but
in a~$3+1$ formulation are arbitrarily specifiable. In the present
setting, their choice is dictated by the explicit Bondi-like gauge
imposed. Adopting the terminology of~\cite{KhoNov02} we can classify
between algebraic and differential gauge choices:
\begin{description}
\item[\textit{Affine null}]It is a complete algebraic gauge for the lapse and
  shift, which is apparent by
  combining~\eqref{eqn:gauge_cond_null_gamma_dd}
  and
  \begin{align*}
    \beta^\rho =  \alpha^2 - 1/f'
    \,,
  \end{align*}
  which results from~$g^{ur}=-1=1/g_{ur}$. The determinant
  condition~\eqref{eqn:gauge_cond_det} does not act as a constraint
  among the three unknown metric components of the two-sphere, but
  merely relates them to the areal radius~$\hat{R}$ that is an
  unknown. The six equations of the main
  system~\eqref{eqn:main_BS_sys} correspond to the six ADM equations
  for~$K_{ij}$ (if~$f' \neq 1$) with a specific addition of
  Hamiltonian and momentum constraints, as well as the lapse and
  shift.
\item[\textit{Bondi-Sachs proper}] This gauge choice is completed by
  the definition of the determinant~\eqref{eqn:gauge_cond_det}. As we
  show in Sec.~\ref{Subsection:BS_proper_gauge} this definition can be
  viewed as providing a differential relation for the shift vector
  component~$\beta^\rho$. In this sense, the Bondi-Sachs gauge proper
  is a mixed algebraic-differential gauge in terms of the lapse and
  shift.
\item[\textit{Double null}] It is also a complete algebraic gauge. The
  complete gauge choice is implied by~$g^{rr}=0$, which combined
  with~$g^{uu}=0$ yields~$\beta^\rho = 0$.
\end{description}

\subsection{Coordinate light speeds}
\label{Subsection:coord_lightspeeds}

Bondi-like gauges are constructed using either incoming or outgoing
null geodesics (or both). It is therefore natural to examine the
coordinate light speeds in these gauges. It is helpful to employ
a~$2+1$ split of the spatial metric~$\gamma_{ij}$ for this purpose. We
briefly review the key elements of this decomposition as necessary for
our discussion. The interested reader can find a complete presentation
in~\cite{Hil15}.

Level sets of constant~$\rho$ are two-spheres. The coordinate~$\rho$
defines an outward pointing normal vector on these spheres
\begin{align}
  s_{(\rho)}^i \equiv \gamma^{ij} L D_j \rho\,,
  \quad
  L^{-2} \equiv \gamma^{ij}(D_i\rho)(D_j\rho)\,.
  \label{eqn:def_s2+1_L}
\end{align}
We call~$L$ the length scalar. The induced metric on two-spheres of
constant~$\rho$ is
\begin{align}
  q_{(\rho)\,ij} \equiv \gamma_{ij} - s_{(\rho)\, i} s_{(\rho)\,j}
  \,,
  \label{eqn:2+1_gamma_split}
\end{align}
where the indices of~$s_{(\rho)}^i$ and~$q_{(\rho)\,ij}$ are lowered
and raised with~$\gamma_{ij}$ and its inverse. Let~$\rho^i$ be the
vector tangent to the lines of constant angular coordinates~$x^A$,
i.e.,~$\rho^i = (\p_\rho)^i$. Then
\begin{align}
  \rho^i = L s_{(\rho)}^i + b^i
  \,,
  \label{eqn:s_rho_1}
\end{align}
where~$b^i s_{(\rho)i} = 0$ and~$b^i$ is called the slip vector. The
length scalar~$L$ and the slip vector~$b^i$ are analogous to the~$3+1$
lapse and shift. They are not, however, freely specifiable but rather
are pieces of the spatial metric~$\gamma_{ij}$.

\begin{figure}[!t]
  \includegraphics[width=0.325\textwidth]{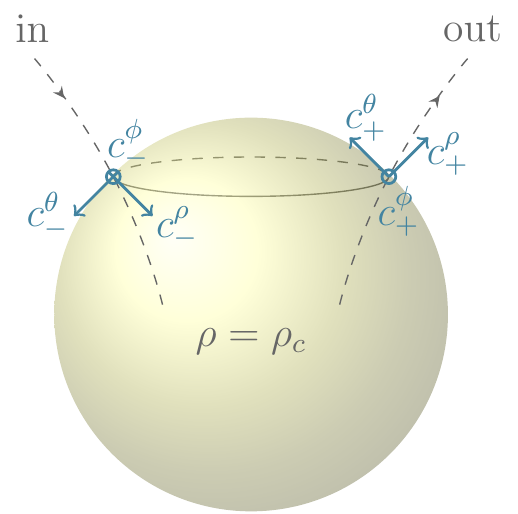}
  \caption{The coordinate light speeds for ingoing and outgoing null
    rays that pass through a surface of constant radius~$\rho_c$,
    i.e., a two-sphere in this example. In an outgoing Bondi-like
    gauge~$c_+^\theta = 0 = c_+^\phi$; i.e., the
    coordinates~$\theta,\phi$ are Lie dragged along the outgoing null
    ray. This ray is orthogonal to the depicted two-sphere.}
  \label{Fig:coord_lightspeeds}
\end{figure}

Let~$\gamma(t)$ be a null curve parametrized by~$t$
and~$L^\mu = \dot{x}^\mu = (1,\dot{x}^i(t))$ a null vector tangent
to~$\gamma(t)$. The coordinate light speeds
are~$C^i \equiv \dot{x}^i(t)$.  Let us further assume that the chosen
null vector obeys the relation
\begin{align}
  L^\mu \propto n^\mu \pm s^\mu_{(\rho)}
  \,.
  \label{eqn:L_mu_1}
\end{align}
From~\eqref{eqn:def_s2+1_L} we get
\begin{align}
  s^\mu_{(\rho)} = (0,L^{-1},L \gamma^{\rho A})
  \,.
  \label{eqn:s_rho_2}
\end{align}
Using~$\rho^\mu = (0,1,0,0)$, solving~\eqref{eqn:s_rho_1}
for~$s^i_{(\rho)}$, and comparing with~\eqref{eqn:s_rho_2}, we obtain
\begin{align}
  b^\rho =0 \,, \quad -b^A = L^2 \gamma^{\rho A}\,.
  \label{eqn:slip_vector_relations}
\end{align}
After multiplying~\eqref{eqn:L_mu_1} with~$\alpha$ we have
\begin{align*}
  L^\mu \propto
  (1, -  \beta^\rho \pm \alpha L^{-1}, -  \beta^A \mp \alpha L^{-1} b^A) 
\end{align*}
from which we read off the coordinate light speeds along null curves
orthogonal to level sets of constant~$\rho$.  They are
\begin{align}
  c_{\pm}^\rho = - \beta^\rho \pm \alpha L^{-1}
  \label{eqn:radial_coord_light_speed}
\end{align}
in the radial direction and
\begin{align}
  c_{\pm}^A = - \beta^A \mp b^A \alpha L^{-1}
  \label{eqn:transverse_coord_light_speed}
\end{align}
in the angular directions. The subscript~$\pm$ refers to
outgoing/ingoing trajectories.  See Fig.~\ref{Fig:coord_lightspeeds}
for an illustration of the coordinate light
speeds. For~$f(\rho)=\rho$, using
Eq.~\eqref{eqn:slip_vector_relations}, the gauge
conditions~\eqref{eqn:gauge_cond_null_gamma_uu} yield
\begin{align}
  c_+^\rho = 1 \,, \quad c_+^A = 0
  \,,
  \label{eqn:bondi-like_coord_light_speeds}
\end{align}
which just expresses the fact that transverse coordinates are Lie
dragged along outgoing null geodesics. For an ingoing single-null
Bondi-like characteristic formulation~$c_+^i \rightarrow c_-^i$ and
for double null~$c_\pm^\rho = \pm 1$. Away from spherical symmetry it
is not generally possible to have~$c_\pm^A$ both vanishing.

\section{Gauge fixing and the principal symbol}
\label{Section:princ_symbol}

Following closely~\cite{KhoNov02,HilRic13} we now discuss the
structure of the principal symbol of the systems we
analyze. See~\cite{AbaReu18,Aba21} for interesting related work on
systems with constraints. As shown in~\cite{HilRic13}, working with
the ADM formalism, in this context, one can distinguish among the
gauge, constraint and physical variables of the system. This
distinction is reflected in the structure of the principal symbol and
allows us to understand which gauges can possibly result in SH
systems.

\paragraph*{FT$2$S Systems and their principal part:} According
to~\cite{GunGar05,HilRic13a} the general first order in time and
second in space (FT$2$S) linear constant coefficient system that
admits a standard first order reduction is of the form
\begin{align}
  \p_t \mathbf{v}
  & = \mathbfcal{A}_1^i \p_i \mathbf{v} + \mathbfcal{A}_1 \mathbf{v}
  + \mathbfcal{A}_2 \mathbf{w} + \mathbf{S}_{\mathbf{v}}
    \,,\nonumber
  \\
  \p_t \mathbf{w}
  & =
  \mathbfcal{B}_1^{ij} \p_i \p_j \mathbf{v}
  + \mathbfcal{B}_1^i \p_i \mathbf{v} + \mathbfcal{B}_1 \mathbf{v}
  + \mathbfcal{B}_2^i \p_i \mathbf{w}\nonumber\\
  &\quad+ \mathbfcal{B}_2 \mathbf{w}
  + \mathbf{S}_{\mathbf{w}}
  \,,\label{FT2S}
\end{align}
where~$\mathbf{S}_{\mathbf{v}}$ and~$\mathbf{S}_{\mathbf{w}}$ are
forcing terms
and~$\mathbfcal{A}_1^i,\, \mathbfcal{A}_2,\,
\mathbfcal{B}_1^{ij},\,\mathbfcal{B}_2^i$ the principal matrices. In
the linear constant coefficient approximation the ADM equations lie in
this category. By \textit{standard first order reduction} we mean one
in which all first order derivatives (temporal and spatial) of
variables that appear with second order derivatives are introduced as
auxiliary variables. We call any first order reduction different from
the aforementioned~\textit{nonstandard}. In such a case only a subset
of the first order derivatives of a variable that appears up to second
order is introduced as auxiliary variables. Or else specific higher
derivatives could be. Given an arbitrary unit spatial covector~$s_i$
(not to be confused with~$s_{(\rho)}^i$ from the previous section),
the principal symbol of the system in the~$s_i$ direction is defined
as
\begin{align}
  \mathbf{P}^s = \left(
  \begin{array}{cc}
    \mathbfcal{A}_1^s & \mathbfcal{A}_2\\
    \mathbfcal{B}_1^{ss} & \mathbfcal{B}_2^s
  \end{array}\right)
  \,,\label{second_order_principal_symbol}
\end{align}
where~$\mathbfcal{A}_1^s\equiv \mathbfcal{A}_1^is_i$ (and so
forth). Writing~$\mathbf{u}=(\p_s\mathbf{v},\mathbf{w})$, we have
\begin{align}
  \p_t\mathbf{u} \simeq \mathbf{P}^s \p_s\mathbf{u}\,,
  \label{eqn:second_order_principal_symbol_eom_form}
\end{align}
where here we dropped nonprincipal terms and all derivatives
transverse to~$s^i$. The definitions of weak and strong hyperbolicity
are identical to those discussed for first order systems in the
Introduction; weak hyperbolicity is the requirement that the
eigenvalues of~$\mathbf{P}^s$ are real for each~$s^i$, and strong
hyperbolicity furthermore is uniformly diagonalizable in~$s^i$. The
second order principal symbol~\eqref{second_order_principal_symbol} is
inherited as a diagonal block of the principal symbol of any standard
first order reduction, where the latter furthermore takes an upper
block triangular form. Consequently only strongly hyperbolic second
order systems may admit a standard first order reduction that is
strongly hyperbolic. The importance of this is that~\eqref{FT2S} has a
well-posed initial value problem in the norm
\begin{align*}
E_1=\sum_i||\partial_i \mathbf{v}||_{L^2}+||\mathbf{w}||_{L^2}\,,
\end{align*}
if and only if it is strongly hyperbolic, where here the norms are
defined over spatial slices of constant~$t$. For our analysis, observe
that the original characteristic form of the equations of motion is
not of the form~\eqref{FT2S}, even after linearization. This issue is
overcome by working instead with the ADM equivalent obtained in
Sec.~\ref{Section:BS_to_ADM}. Working with the equivalent furthermore
has the advantage that the theory discussed below was developed in
this language, making application straightforward. Due to the freedom
in choosing a time slicing, there is freedom in the construction of
the equivalent ADM formulation. This was parametrized by~$f'(\rho)$ in
the previous section. For brevity we work assuming~$f'(\rho)=1$, but
since the structural properties discussed above hold true in {\it any}
alternative slicing, this restriction does not affect the outcome of
the analysis.

\paragraph*{Pure gauge degrees of freedom:} In many cases of physical
interest~FT$2$S systems arise with additional structure in their
principal symbol. In GR for instance, structure arises as a
consequence of gauge freedom. To see this, suppose that we are working
in a coordinate basis with an arbitrary solution to the vacuum field
equations. The field equations are, of course, invariant under changes
of coordinates~$x^{\mu}\to X^{\underline{\mu}}$, so that both the
metric and curvature transform in the same manner. This invariance has
important consequences on the form of the field equations. Consider an
infinitesimal change to the coordinates
by~$x^{\mu}\to x^{\mu}+\xi^\mu$. Such a change results in a
perturbation to the metric of the form
\begin{align*}
\delta g_{\mu\nu}= -\nabla_\mu \xi_\nu - \nabla_\nu \xi_\mu = -\Lie_\xi g_{\mu\nu}\,.
\end{align*}
This transformation, the linearization of the condition for covariance
in a coordinate basis, simultaneously serves as the gauge freedom of
linearized GR. Working now in the ADM language, and~$3+1$
decomposing~$\xi^a$ by
\begin{align*}
  \Theta \equiv - n_\mu \xi^\mu
  \,, \quad
  \psi^i \equiv - \gamma^i {}_\mu \xi^\mu
  \,, 
\end{align*}
the pure gauge perturbations~$(\Theta,\psi^i)$ satisfy,
\begin{align}
\p_t\Theta&=\delta\alpha-\psi^iD_i\alpha+\Lie_\beta\Theta\,,\nonumber\\
\p_t\psi^i&=\delta\beta^i+\alpha D^i\Theta-\Theta D^i\alpha+\Lie_\beta\psi^i\,,
\label{eqn:PG_evolution}
\end{align}
with~$\delta\alpha$ and~$\delta\beta^i$ the perturbation of the lapse
and shift respectively. The resulting perturbation to the metric and
extrinsic curvature can be explicitly computed~\cite{HilRic13}, and
are given by,
\begin{subequations}
  \label{eqn:gauge_transf_2}
  \begin{align}
    \delta \gamma_{ij}
    &= - 2 \Theta K_{ij} + \mathcal{L}_\psi \gamma_{ij}
    \,,
    \label{eqn:delta_gamma_gauge_tranf}
    \\
    \delta K_{ij}
    & =
    - D_i D_j \Theta
    \nonumber \\
    & \quad
    + \Theta \big( R_{ij} - 2 K^k {}_i K_{jk} + K_{ij} K \big)
    + \mathcal{L}_\psi K_{ij}
    \,,
    \label{eqn:delta_K_gauge_tranf}
  \end{align}
\end{subequations}
where~$\gamma_{ij}$ and~$K_{ij}$ are to be understood by their
background values. It is a remarkable fact that these equations are
nothing more than the ADM evolution equations under the
replacements~$\alpha\to\Theta$ and~$\beta^i\to\psi^i$, so that the ADM
evolution equations can be interpreted as a local gauge transformation
in a coordinate basis. Given a choice for either the lapse and shift,
or an equation of motion for each, or a combination thereof, we may
combine~\eqref{eqn:PG_evolution} and~\eqref{eqn:gauge_transf_2}, to
obtain a closed system for the pure gauge variables~$(\Theta,\psi^i)$
and~$(\delta\alpha,\delta\psi^i),$ on the background spacetime. We
call this the pure gauge subsystem. Suppose, for example, that we
employed a harmonic time coordinate~($\Box t=0$) with a vanishing
shift. In~$3+1$ language this gives
\begin{align*}
\p_t\alpha&=-\alpha^2K.
\end{align*}
The pure gauge subsystem~\eqref{eqn:PG_evolution}
for~$(\Theta,\psi^i)$ is then completed by
\begin{align*}
  \p_t\delta\alpha&\simeq \alpha^2\p^i\p_i\Theta\,,\qquad
  \delta\beta^i=0\,,
\end{align*}
where we have used~\eqref{eqn:gauge_transf_2} and discarded
nonprincipal terms. The additional structure alluded to above is that
for a given choice of gauge, the principal symbol of the pure gauge
subsystem is inherited as a sub-block of the principal symbol of any
formulation of GR that employs said gauge. This is demonstrated by
using suitable projection operators which are stated explicitly below.

\paragraph*{Constraint violating degrees of freedom:} Yet more
structure arises from the constraints. Assuming the ADM evolution
equations hold, the Hamiltonian and momentum constraints formally
satisfy evolution equations,
\begin{align*}
  \p_tH&=-2\alpha D^iM_i-4M^iD_i\alpha+2\alpha K H+\Lie_\beta H\,,\nonumber\\
  \p_tM_i&= -\tfrac{1}{2}\alpha D_iH+\alpha K M_i-D_i\alpha H+\Lie_\beta M_i\,,
\end{align*}
so that constraints satisfying initial data remain so in their domain
of dependence. These equations follow from the contracted Bianchi
identities. In free-evolution formulations of GR however, the ADM
evolution equations need not hold, since combinations of the
constraints can be freely added to the evolution equations. Doing so
results in adjusted evolution equations for the constraints, which
nevertheless remain a closed set of equations. Just as the principal
symbol of the full equations of motion inherit the pure gauge
principal symbol, the principal symbol of the constraint subsystem
manifests as a sub-block. This is again seen using the projection
operators stated below.

\paragraph*{Linearized ADM:} To apply straightforwardly the theory described
at the start of this section we linearize about flat space in global
inertial coordinates. The analysis can be carried out around a general
background leading to the same conclusions. In this setting we obtain
for the metric and extrinsic curvature perturbations the evolution
equations,
\begin{subequations}
  \label{eqn:ADM_lin_mink}
  \begin{align}
    \p_t \delta \gamma_{ij}
    & = -2 \delta K_{ij} + \p_{(i} \delta \beta_{j)}
      \,,
      \label{eqn:ADM_lin_mink_gamma}
    \\
    \p_t \delta K_{ij}
    & =
      - \p_i \p_j \delta \alpha
      - \tfrac{1}{2}\p^k \p_k \delta \gamma_{ij}
      \nonumber \\
    & \quad
      -\tfrac{1}{2} \p_i \p_j \delta \gamma
      + \p^k \p_{(i} \delta \gamma_{j)k}
      \,.
      \label{eqn:ADM_lin_mink_K}
  \end{align}
\end{subequations}
The constraints become
\begin{align*}
  \delta H&=\p^i\p^j\delta\gamma_{ij}-\p^i\p_i\delta\gamma \,,\nonumber\\
  \delta M_i&=\p^j\delta K_{ij}-\p_i \delta K  \,,
\end{align*}
and evolve according to
\begin{align}
  \label{eqn:Cons_subsys_lin_mink}
  \begin{aligned}
  \p_t\delta H&= -2 \p^i\delta M_i\,,\\
  \p_t\delta M_i&= -\tfrac{1}{2} \p_i\delta H\,.
  \end{aligned}
\end{align}
About this background the pure gauge
equations~\eqref{eqn:PG_evolution} simplify to
\begin{subequations}
  \label{eqn:PG_subsys_lin_mink}
  \begin{align}
  \p_t \Theta
   & = \delta \alpha
     \,,
     \label{eqn:PG_subsys_lin_mink_theta}
  \\
  \p_t \psi_i
   &=
     \delta \beta_i + \p_i \Theta
     \,.
     \label{eqn:PG_subsys_lin_mink_psi}
  \end{align}
\end{subequations}

\paragraph*{Pure gauge projection operators:} Let~$s^i$ be an arbitrary
constant spatial unit vector. To extract the gauge, constraint and
physical degrees of freedom within the principal symbol in this
direction we must decompose the state vector appropriately. The
induced metric on the surface transverse to~$s^i$ is
\begin{align*}
  q_{ij} \equiv \gamma_{ij} - s_i s_j \,.
\end{align*}
Here we denote by~$\hat{A},\,\hat{B}$ the spatial directions
transverse to~$s^i$, which--since in
general~$s^i \neq s_{(\rho)}^i$--do not necessarily coincide with the
angular directions from our earlier discussion. Projections of the ADM
variables that capture pure gauge equations of
motion~\eqref{eqn:PG_subsys_lin_mink} are given by
\begin{align}
  \label{eqn:gauge_vars}
  \begin{aligned}
    [\p_s^2 \Theta]
    &= - \delta K_{ss}
    \,, \quad
    [\p_s^2 \psi_s] = \tfrac{1}{2} \p_s \delta \gamma_{ss}
    \\
    [\p_s^2 \psi_{\hat{A}}]
    &= \p_s \delta \gamma_{s \hat{A}}
    \,.
  \end{aligned}
\end{align}
Here the notation~$[\cdots]$ is used to emphasize that the specific
projection of the ADM variables on the right-hand side shares, within
the principal symbol, the structure of the pure gauge variable named
on the left-hand side. This is spelled out below. Thus, together
with~$\p_s \delta \alpha,\, \p_s \delta \beta_s,\, \p_s \delta
\beta_{\hat{A}}$ they encode the complete pure gauge variables of the
system, with~$\delta \alpha,\, \delta \beta^i$ the perturbation to the
lapse and shift.

\paragraph*{Constraint projection operators:} Likewise, within the
principal symbol the Hamiltonian and momentum constraints are encoded
by the projections,
\begin{align}
  \begin{aligned}
    [H]&=-\p_s\delta\gamma_{qq}\,,\quad [M_s]=-\delta K_{qq}\,,\\
    [M_{\hat{A}}]&=\delta K_{s\hat{A}}\,,
  \end{aligned}
  \label{eqn:constr_vars}
\end{align}
with the naming convention as above. Here and in the following
indices~$qq$ denote that the trace was taken with~$q^{ij}$.

\paragraph*{Physical projection operators:} Finally, the remaining
variables to be taken account of are the trace-free
projections. Defining the projection operator familiar from textbook
treatments of linear gravitational waves,
\begin{align}
  {}^P \! \! \! \perp^{kl}_{ij}
  \equiv
  q^k_{\;(i} q^l{}_{j)} -  \tfrac{1}{2} q_{ij} q^{kl}
  \,.
  \label{eqn:phys_projector}
\end{align}
we define 
\begin{align*}
  \begin{aligned}
    \p_s\delta \gamma_{\hat{A} \hat{B}}^{\textrm{TF}}
    &={}^P \! \! \! \perp^{ij}_{\hat{A}\hat{B}} \p_s\delta \gamma_{ij} 
    \,,\quad
    \delta K_{\hat{A} \hat{B}}^{\textrm{TF}}
    ={}^P \! \! \! \perp^{ij}_{\hat{A}\hat{B}} \delta K_{ij} 
    \,.
  \end{aligned}
\end{align*}
The superscript~$\textrm{TF}$ denotes trace-free. These variables are
associated with the physical degrees of freedom.

\paragraph*{The principal symbol:} Employing the notation above we can now
write out the principal symbol in the
form~\eqref{eqn:second_order_principal_symbol_eom_form}. Starting with
the pure gauge block, this gives
\begin{align}
  \label{eqn:PG_Ps}
  \begin{aligned}
    \p_t[\p_s^2\Theta] &\simeq \p_s(\p_s\delta\alpha) + \tfrac{1}{2}\p_s[H]\,,\\
    \p_t[\p_s^2\psi_s] &\simeq \p_s(\p_s\delta\beta_s) + \p_s[\p_s^2\Theta]\,,\\
    \p_t[\p_s^2\psi_{\hat{A}}] &\simeq \p_s(\p_s\delta\beta_{\hat{A}})
    -2\p_s[M_{\hat{A}}]\,.
  \end{aligned}
\end{align}
Comparing this with~\eqref{eqn:PG_subsys_lin_mink} it is clear that up
to additions of the ``constraint variables'' there is agreement. Next,
the constraint violating block gives
\begin{align}
  \label{eqn:Cons_Ps}
  \begin{aligned}
    \p_t[H] &\simeq -2\p_s [M_s] \,,\\
    \p_t[M_s] &\simeq - \tfrac{1}{2}\p_s[H]\,,\\
    \p_t[M_{\hat{A}}] &\simeq 0\,.
  \end{aligned}
\end{align}
Comparing this with~\eqref{eqn:Cons_subsys_lin_mink} there is perfect
agreement. Finally the physical block is
\begin{align}
  \label{eqn:physical_subsys}
  \begin{aligned}
    \p_t\p_s\delta \gamma_{\hat{A} \hat{B}}^{\textrm{TF}}
    &\simeq -2 \p_s\delta K_{\hat{A} \hat{B}}^{\textrm{TF}}
    \,,\\
    \p_t \delta K_{\hat{A} \hat{B}}^{\textrm{TF}}
    & \simeq
    - \tfrac{1}{2} \p_s^2\delta \gamma_{\hat{A} \hat{B}}^{\textrm{TF}}
    \,,
  \end{aligned}
\end{align}
which is decoupled from the rest of the equations. These equations are
not yet complete, because we have not yet made a concrete choice of
gauge. Several Bondi-like gauges are treated in detail in the
following sections.

\paragraph*{Discussion:} The results of the foregoing discussion follow
because GR is a constrained Hamiltonian system that satisfies the
hypotheses of~\cite{HilRic13}. To make the presentation here somewhat
more stand-alone however, let us consider a plane wave ansatz
\begin{align}
  \label{eqn:Met_Decomp}
  \begin{aligned}
    \delta\gamma_{ij} &= 2 e^{\kappa^{(\psi_s)}_\mu x^\mu} s_is_j[\p_s\tilde{\psi_s}]
    - \tfrac{1}{2} q_{ij} e^{\kappa^{(H)}_\mu x^\mu} [\tilde{H}]\\
    &+2e^{\kappa^{(\psi_A)}_\mu x^\mu}q^{\hat{A}}{}_{(i}s_{j)}[\p_s\tilde{\psi}_{\hat{A}}]
    + e^{\kappa^{(P)}_\mu x^\mu}
    {}^P \! \! \! \perp^{\hat{A}\hat{B}}_{ij}
    \delta\gamma_{\hat{A} \hat{B}}^{\textrm{TF}} \,,\\
    \delta K_{ij} &=  - e^{\kappa^{(\Theta)}_\mu x^\mu} s_is_j[\p^2_s\tilde{\Theta}]
    - \tfrac{1}{2} q_{ij} e^{\kappa^{(M_s)}_\mu x^\mu} [\tilde{M}_s]\\
    &+2e^{\kappa^{(M_A)}_\mu x^\mu}q^{\hat{A}}{}_{(i}s_{j)}[\tilde{M}_{\hat{A}}]
    + e^{\kappa^{(P)}_\mu x^\mu}
    {}^P \! \! \! \perp^{\hat{A}\hat{B}}_{ij} \delta K_{\hat{A} \hat{B}}^{\textrm{TF}}\,,
  \end{aligned}
\end{align}
with each wave vector of the
form~$\kappa_\mu = (\kappa, i\,\omega s_i)$. These solutions travel in
the~$\pm s^i$ directions, although since the lapse and shift are as
yet undetermined, the~$\kappa$'s cannot be solved for so far. Defining
the projections exactly as above, the unknowns can be decomposed {\it
  explicitly} into their gauge, constraint violating and gravitational
wave pieces as indicated by the naming, and
Eqs.~\eqref{eqn:PG_Ps},~\eqref{eqn:Cons_Ps},
and~\eqref{eqn:physical_subsys} become exact. In the nonlinear setting
it is, of course, hopeless to try and decompose metric components into
constituent gauge, constraint violating and physical degrees of
freedom. But even in the linear constant coefficient approximation,
solutions consist in general of a sum over many such plane waves
propagating in different directions, and so the
decomposition~\eqref{eqn:Met_Decomp} is not a sufficient description.
What is important for our purposes however, is that the structure in
the field equations that permits the
decomposition~\eqref{eqn:Met_Decomp} for plane wave solutions is
present regardless of the direction~$s^i$ considered. The principal
symbol sees only this structure and thus, with
Eqs.~\eqref{eqn:PG_Ps},~\eqref{eqn:Cons_Ps}
and~\eqref{eqn:physical_subsys} above completed with a choice for the
lapse and shift, can be written in the schematic form
\begin{align}
  \mathbf{P}^s =
  \begin{pmatrix}
    \mathbf{P}_G & \mathbf{P}_{GP} & \mathbf{P}_{GC} \\
    0 & \mathbf{P}_P & \mathbf{P}_{PC} \\
    0 & 0 &  \mathbf{P}_C
  \end{pmatrix}
  \,,\label{eqn:princ_symbol_triang}
\end{align}
even upon linearization about an arbitrary
background. Here~$\mathbf{P}_G,\,\mathbf{P}_C,\,\mathbf{P}_P$ denote
the gauge, constraint and physical sub-blocks
and~$\mathbf{P}_{GC},\,\mathbf{P}_{GP},\,\mathbf{P}_{PC}$ parametrize
the coupling between them. As seen in~\cite{HilRic13} there is a very
large class of gauge conditions and natural constraint additions that
result
in~$\mathbf{P}_{GP}=\mathbf{P}_{GC}=\mathbf{P}_{PC}=0$. Consequently,
it follows from~\eqref{eqn:princ_symbol_triang} that a necessary
condition for strong hyperbolicity of the formulation is that the pure
gauge and constraint subsystems are themselves strongly
hyperbolic. Following~\cite{KhoNov02} we may therefore restrict our
attention first to pure gauge systems of interest, which have the
advantage of being smaller, and thus are much easier to treat.

\paragraph*{Bondi-like gauges:} The gauges we are concerned with all require
the condition~\eqref{eqn:gauge_cond_null}, which in characteristic
coordinates implies the same for the perturbation to the metric, that
is,
\begin{align*}
  \delta g^{uu} = \delta g^{uA} = 0 \,.
\end{align*}
There remains one gauge condition to be specified, namely the
parametrization along outgoing null surfaces by a radial coordinate.
Next we study specific instances of this condition.

\section{The affine null gauge}
\label{Section:aff_null_gauge}

In this section we analyze the degree of hyperbolicity of the EFE in
the affine null gauge~\cite{Win13, CreOliWin19}. In~\cite{GiaHilZil20}
a hyperbolicity analysis for the EFE in the affine null gauge for
asymptotically AdS five-dimensional spacetime with planar symmetry was
performed and the full system was shown to be WH. Here, we do this
analysis in four-dimensional asymptotically flat spacetime, but more
importantly we also analyze the pure gauge subsystem and show that the
weak hyperbolicity of the full system stems from that of the pure
gauge subsystem.

The complete affine null gauge fixing is given by
\begin{align}
  \alpha = L^{-1}
  \,, \quad
  \beta^\rho = L^{-2} -1
  \,, \quad
  \beta^A = - b^A L^{-2}
  \,,
  \label{eqn:aff_null_gauge}
\end{align}
where Eqs.~\eqref{eqn:radial_coord_light_speed}
-\eqref{eqn:bondi-like_coord_light_speeds} and~$g^{ur}=-1$ have been
combined. As in the previous section, we work in the linear, constant
coefficient approximation, and for simplicity we assume that
in~\eqref{eqn:coord_transf}~$f(\rho)=\rho$.

\subsection{Pure gauge subsystem}
\label{Subsection:aff_null_gauge:pg_subsys}

Let us first consider pure gauge metric
perturbations~\eqref{eqn:gauge_transf_2}. To close the
system~\eqref{eqn:PG_subsys_lin_mink} further input for~$\delta
\alpha$ and~$\delta \beta^i$ is needed. For the affine null gauge this
follows from~\eqref{eqn:aff_null_gauge}, which after linearization
about flat space reads
\begin{align*}
  \delta \alpha
  &= -\tfrac{1}{2} \delta \gamma_{\rho \rho}
    \,, \quad \;  
    \delta \beta^\theta
    = - \rho^{-2} \delta \gamma_{\rho \theta}
    \,,
  \\
  \delta \beta^\rho
  &= - \delta \gamma_{\rho \rho}
    \,, \qquad
    \delta \beta^{\phi} = - \rho^{-2} \sin^2\theta \,
    \delta \gamma_{\rho \phi}
    \,.
    \nonumber
\end{align*}
Using~$\delta \gamma_{ij} =  \p_i \psi_j + \p_j \psi_i$
and~$\psi^i = \gamma^{ij} \psi_j$ the latter reads
\begin{align}
  \delta \alpha
  &= - \p_\rho \psi^\rho
    \,, \quad \, 
  \delta \beta^\theta
  = - \p_\rho \psi^\theta - \rho^{-2} \p_\theta \psi^\rho
    \,,
    \label{eqn:aff_null_coord_pert}
  \\
  \delta \beta^\rho
  &= - 2 \p_\rho \psi^\rho
    \,, \; \;
    \delta \beta^{\phi} =
    -\p_\rho \psi^\phi  - \rho^{-2} \sin^2\theta \, \p_\phi \psi^\rho
    \,.
    \nonumber
\end{align}
The pure gauge subsystem~\eqref{eqn:PG_subsys_lin_mink} is then
\begin{subequations}
  \label{eqn:aff_null_pg}
  \begin{align}
    &
    (\p_t+\p_\rho) ( \psi^\rho - \Theta) =  0
      \,,
      \label{eqn:aff_null_pg_a}
    \\
    &
    (\p_t + \p_\rho) \psi^\rho + \p_\rho(\psi^\rho - \Theta) =  0
      \,,
      \label{eqn:aff_null_pg_b}
    \\
    &
    (\p_t + \p_\rho) \psi^\theta
      + \rho^{-2} \p_\theta (\psi^\rho - \Theta) =  0
      \,,
      \label{eqn:aff_null_pg_c}
     \\
    &
    (\p_t + \p_\rho) \psi^\phi
    + (\rho \sin \theta)^{-2} \p_\phi (\psi^\rho - \Theta) =  0
      \,,
      \label{eqn:aff_null_pg_d}
  \end{align}
\end{subequations}
where~$\p_t + \p_\rho =\p_r$ is an outgoing null derivative
and~\eqref{eqn:aff_null_pg_a} results from a linear combination
of~\eqref{eqn:PG_subsys_lin_mink_theta}
and~\eqref{eqn:PG_subsys_lin_mink_psi} with~$i=\rho$. Along an
arbitrary spatial direction~$s^i$ a first order linear system can be
written as
\begin{align*}
  \p_t \mathbf{v} \simeq \mathbf{J}^s \p_s \mathbf{v}
  \,,
\end{align*}
where~$\mathbf{T}^{-1}_s \mathbf{P}^s \mathbf{T}_s \equiv
\mathbf{J}^s$ is the Jordan normal form of the principal
symbol,~$\mathbf{P}^s$,~$\mathbf{v} \equiv \mathbf{T}^{-1}_s
\mathbf{u}$ are the associated (generalized) characteristic variables
and~$\simeq$ denotes equality up to source terms and derivatives
transverse to~$s^i$. The principal symbol of the pure gauge
subsystem~\eqref{eqn:aff_null_pg} is clearly nondiagonalizable along
the~$\rho,\theta, \phi$ directions, and, in fact, in any
direction. In~\eqref{eqn:aff_null_pg_b},~\eqref{eqn:aff_null_pg_c},
and~\eqref{eqn:aff_null_pg_d} the
terms~$\p_\rho(\psi^\rho - \Theta)$,~$\p_\theta(\psi^\rho - \Theta)$,
and~$\p_\phi(\psi^\rho - \Theta)$ result in~$2 \times 2$ Jordan
blocks, along~$\rho,\theta$, and~$\phi$, respectively. The principal
symbol of the full set of equations of motion for GR has the upper
triangular form~\eqref{eqn:princ_symbol_triang} when a standard first
order reduction is considered. Thus it will possess nontrivial Jordan
blocks along all~$\rho,\theta,\phi$ directions as well. In
Secs.~\ref{Subsec:aff_null:pg_sub-block:rho_dir}
and~\ref{Subsec:aff_null:pg_sub-block:theta_dir} we show this
explicitly and demonstrate the connection to the PDE system in
characteristic coordinates.

An intriguing observation is that the pure gauge
variable~$(\psi^\rho-\Theta)$ satisfies a transport equation
along~$\p_r$. So, acting from the left on~\eqref{eqn:aff_null_pg}
with~$\p_r$ and commuting the spatial and null derivatives
on~$(\psi^\rho - \Theta)$, one obtains
\begin{subequations}
  \begin{align}
    &
    \p_r^2 (\Theta - \psi^\rho)  = 0
      \,,
      \label{eqn:aff_null_pg_2nd_order_a}
    \\
    &
    \p_r^2 \psi^\rho   = 0
      \,,
      \label{eqn:aff_null_pg_2nd_order_b}
    \\
    &
    \p_r^2 \psi^\theta
      =  0
      \,,
      \label{eqn:aff_null_pg_2nd_order_c}
     \\
    &
    \p_r^2 \psi^\phi
      = 0
      \,.
      \label{eqn:aff_null_pg_2nd_order_d}
  \end{align}
  \label{eqn:aff_null_pg_2nd_order}%
\end{subequations}
This system admits a nonstandard reduction to first order which is
strongly hyperbolic. To see this, we introduce only outgoing null
derivatives of the unknowns as auxiliary variables. All of the
variables then satisfy transport equations in the outgoing null
direction. In contrast to this, for a standard first order reduction
both the time and space derivatives of the unknowns would be
introduced as auxiliary variables.

The relevant question is whether there exists a formulation of GR that
inherits the structure of the second version of the pure gauge
subsystem~\eqref{eqn:aff_null_pg_2nd_order}, rather than the
first~\eqref{eqn:aff_null_pg}. In view of the results
of~\cite{HilRic13}, if such a formulation exists, it would necessarily
admit a nonstandard first order reduction. In
Sec.~\ref{Subsec:aff_null:pg_sub-block:rho_dir} we show that there is
a convenient combination of ADM variables that allows one to remove
the nontrivial Jordan block along the~$\rho$ direction that appears in
a standard first order reduction. This is true due to the specific
gauge choice and its construction upon outgoing null
geodesics. Crucially, however, this special combination is only
possible along the~$\rho$ direction but not~$\theta,\phi$. So, away
from spherical symmetry the EFE in the affine null gauge are only WH.

\subsection{Pure gauge sub-block: Radial direction}
\label{Subsec:aff_null:pg_sub-block:rho_dir}

We now demonstrate how the radial part of the pure gauge
subsystem~\eqref{eqn:aff_null_pg} is inherited by the linearized
EFE. For brevity in this subsection we work in spherical symmetry,
which is sufficient, since the coupled gauge variables in the radial
Jordan block of~\eqref{eqn:aff_null_pg} are present already under this
assumption.

\subsubsection{ADM setup}

In spherical symmetry the principal part of the linearized ADM
equations in outgoing affine null gauge is
\begin{subequations}
  \label{eqn:ADM_lin_mink_aff_null_spher}
  \begin{align}
  \p_t \delta \gamma_{\rho \rho}
  &\simeq -2 \delta K_{\rho \rho}
    - 2 \p_\rho \delta \gamma_{\rho \rho}
    \,,
    \label{eqn:ADM_lin_mink_aff_null_spher_a}
  \\
  \p_t \delta K_{\rho \rho}
  & \simeq
  \tfrac{1}{2} \p_\rho^2 \delta \gamma_{\rho \rho}
  - \rho^{-2} \p_\rho^2
    \delta \gamma_{\theta \theta}
    \,,
    \label{eqn:ADM_lin_mink_aff_null_spher_b}
  \\
  \p_t \delta \gamma_{\theta \theta}
  & \simeq -2 \delta K_{\theta \theta}
    \,,
    \label{eqn:ADM_lin_mink_aff_null_spher_c}
  \\
  \p_t \delta K_{\theta \theta}
  & \simeq
  - \tfrac{1}{2} \p_\rho^2 \delta \gamma_{\theta \theta}
    \,.
    \label{eqn:ADM_lin_mink_aff_null_spher_d}
\end{align}
\end{subequations}
From Eq.~\eqref{eqn:gauge_vars}, the gauge variables along the~$\rho$
direction in spherical symmetry are
\begin{align}
  &
    -\delta K_{\rho \rho} = [\p_\rho^2 \Theta]
  \,,
  \qquad 
    \tfrac{1}{2} \p_\rho \delta \gamma_{\rho \rho}
    = [\p_\rho^2 \psi^\rho]
    \,.
    \label{eqn:gauge_ADM_vars_spher}
\end{align}
To recover the pure gauge structure it suffices to analyze the
coupling between~\eqref{eqn:ADM_lin_mink_aff_null_spher_a}
and~\eqref{eqn:ADM_lin_mink_aff_null_spher_b},
\begin{subequations}
\label{eqn:ADM_lin_mink_aff_null_spher_2}
\begin{align}
 \p_r (\tfrac{1}{2} \delta \gamma_{\rho \rho})
  & \simeq
    - \delta K_{\rho \rho}
    -  \p_\rho (\tfrac{1}{2}\delta \gamma_{\rho \rho})
    \,,
    \label{eqn:ADM_lin_mink_aff_null_spher_2_a}
  \\
    \label{eqn:ADM_lin_mink_aff_null_spher_2_b}
 \p_r ( \delta K_{\rho \rho}
    + \tfrac{1}{2} \p_\rho \delta \gamma_{\rho \rho})
  & \simeq - \rho^{-2} \p_\rho^2 \delta \gamma_{\theta \theta}
    \,,
\end{align}
\end{subequations}
where~$\p_r=\p_t+\p_\rho$ is an outgoing null vector
and~\eqref{eqn:ADM_lin_mink_aff_null_spher_2_b} results from a linear
combination of~\eqref{eqn:ADM_lin_mink_aff_null_spher_a}
and~\eqref{eqn:ADM_lin_mink_aff_null_spher_b}. The right-hand side
of~\eqref{eqn:ADM_lin_mink_aff_null_spher_2_b} involves the constraint
variable
\begin{align*}
  [H] = - \p_\rho \delta \gamma_{qq}
  = - 2\rho^{-2} \p_\rho
  \delta \gamma_{\theta \theta}\,.
\end{align*}
In a standard first order reduction, the term~$ (\p_\rho \delta
\gamma_{\rho \rho}) $ would be introduced as an evolved variable
satisfying
\begin{align}
  \p_r  (\tfrac{1}{2}\p_\rho \delta \gamma_{\rho \rho})
  & \simeq
    - \p_\rho \delta K_{\rho \rho}
    - \p_\rho (\tfrac{1}{2}\p_\rho \delta \gamma_{\rho \rho})
    \,.
    \label{eqn:ADM_lin_mink_aff_null_spher_3}  
\end{align}
The above and~\eqref{eqn:ADM_lin_mink_aff_null_spher_2_b} expressed in
terms of gauge and constraint variables read
\begin{align*}
  & \p_r [\p_\rho^2 \psi^\rho]
  + \p_\rho [ \p_\rho^2 \left( \psi^\rho - \Theta \right)]
    \simeq 0
    \,,
    \nonumber
  \\
  & \p_r
    [\p_\rho^2 \left(\Theta - \psi^\rho \right)]
    \simeq \tfrac{1}{2} \p_\rho [H]
    \,.
\end{align*}
As explained in Sec.~\ref{Section:princ_symbol}, this system has a
pure gauge part that consists of the coupling among the gauge
variables~$\Theta$ and~$\psi^\rho$ and a part that captures the
coupling of the gauge to the constraint variables. The pure gauge
part~$\mathbf{P}_G$ is obtained by neglecting the
term~$\p_\rho [H]/2$. This part has the same principal structure as
the pure gauge subsystem~\eqref{eqn:aff_null_pg} in the radial
direction, since it is just an overall~$\p_\rho^2$ derivative of the
latter. This is in accordance with the result of~\cite{HilRic13},
because for a standard first order reduction~$\mathbf{P}_G$ inherits
the structure of the first order system formed
by~$(\Theta, \psi_i,\delta \alpha,\delta \beta_i)$. The
term~$\p_\rho [H]/2$ is encoded in the~$\mathbf{P}_{GC}$ sub-block of
the full principal symbol~$\mathbf{P}^\rho$.

Next, let us consider a reduction in
which~$(\p_r \delta \gamma_{\rho \rho})$ is introduced as an auxiliary
variable rather than~$(\p_\rho \delta \gamma_{\rho
  \rho})$. From~\eqref{eqn:ADM_lin_mink_aff_null_spher_2_a}
and~\eqref{eqn:gauge_ADM_vars_spher} we get
\begin{align}
  \p_r (\tfrac{1}{2}\delta \gamma_{\rho \rho})
  =[ \p_r \p_\rho  \psi^\rho]
  \simeq [\p_\rho^2 (\Theta- \psi^\rho)]\,,
  \label{eqn:gauge_ADM_vars_spher_null_dev_a}
\end{align}
where in the first step we are just using our normal naming convention
with~$[\dots]$ and likewise in the second
Eq.~\eqref{eqn:gauge_ADM_vars_spher}. Similarly, from
Eq.~\eqref{eqn:gauge_ADM_vars_spher} we get
\begin{align}
  \tfrac{1}{2} \p_\rho \delta \gamma_{\rho \rho}
  + \delta K_{\rho \rho}
  = [\p_\rho^2(\psi^\rho - \Theta)] =
  [\p_r \p_\rho  \psi^\rho] =
  - [\p_r \p_\rho  \Theta]
  \,,
  \label{eqn:gauge_ADM_vars_spher_null_dev_b}
\end{align}
where in the second step Eq.~\eqref{eqn:aff_null_pg_b} and in the
third Eq.~\eqref{eqn:aff_null_pg_a} are used. The equation of motion
for the auxiliary variable~$(\p_r\delta \gamma_{\rho\rho})$ results
from~\eqref{eqn:ADM_lin_mink_aff_null_spher_2_a} after acting
with~$\p_r$, namely
\begin{align}
\p_r (\tfrac{1}{2}\p_r\delta \gamma_{\rho \rho})
  &
  \simeq - \p_r ( \delta K_{\rho \rho}
  + \tfrac{1}{2} \p_\rho \delta \gamma_{\rho \rho})
    \nonumber
  \\
  & 
    \simeq
    \rho^{-2} \p_\rho^2 \delta \gamma_{\theta \theta}
    \,,
    \label{eqn:p_r_delta_gamma_rhorho_eom}
\end{align}
where in the second step
Eq.~\eqref{eqn:ADM_lin_mink_aff_null_spher_2_b} is used. The above
together with~Eq.~\eqref{eqn:ADM_lin_mink_aff_null_spher_2_b} in terms
of the gauge and constraint variables read
\begin{subequations}
  \label{eqn:pure_gauge_sub-block_2nd_order_spher}
  \begin{align}
     \p_r [  \p_r \p_\rho \psi^\rho]
    &
      \simeq - \tfrac{1}{2} \p_\rho [H]
  \,,
  \label{eqn:pure_gauge_sub-block_2nd_order_spher_a}
  \\
    \p_r [ \p_r \p_\rho  \Theta]
    &
      \simeq \tfrac{1}{2} \p_\rho [H]
  \,,
  \label{eqn:pure_gauge_sub-block_2nd_order_spher_b}
\end{align}
\end{subequations}
where the relations~\eqref{eqn:gauge_ADM_vars_spher_null_dev_a}
and~\eqref{eqn:gauge_ADM_vars_spher_null_dev_b} have been used. Thus,
the system~\eqref{eqn:ADM_lin_mink_aff_null_spher_2_b}
and~\eqref{eqn:p_r_delta_gamma_rhorho_eom} inherits the principal
structure of~\eqref{eqn:aff_null_pg_2nd_order_a}
and~\eqref{eqn:aff_null_pg_2nd_order_b} in~$\mathbf{P}_G$. Again the
term~$\p_\rho [H]/2$ is in the~$\mathbf{P}_{GC}$ sub-block. This
result does not contradict~\cite{HilRic13} due to the nonstandard
first order reduction considered. In the outgoing affine null gauge
the outgoing null direction possesses a special role as the
foundational piece of the construction. This construction provides the
opportunity to group ADM variables in such a way that we can avoid the
nontrivial Jordan block in the radial direction.

\subsubsection{Characteristic setup}

The ADM analysis above teaches us which variables inherit the
principal structure of the pure gauge degrees of freedom. However, the
original PDE problem is formulated in the characteristic
domain. In~\cite{HilRic13} the pure gauge structure was identified for
a spacelike foliation. Whether this is possible in the characteristic
domain is closely related to the existence of the previous first order
reductions in this domain as well. We show here that both previous
first order reductions and their principal structure can be realized
in the characteristic setup directly.

To demonstrate this consider the affine null gauge in an outgoing
characteristic formulation. The complete calculation can be found in
the ancillary files. We first employ the metric ansatz
\begin{align*}
  ds^2 = g_{uu} du^2 -2dudr + g_{\theta \theta} d\theta^2 +  g_{\phi \phi} d\phi^2
  \,,
\end{align*}
which for flat space reads
\begin{align*}
  g_{uu}=-1\,,
  \quad
  g_{\theta \theta}=r^2\,,
  \quad
  g_{\phi \phi}=r^2 \sin^2 \theta
  \,.
\end{align*}
Analyzing the main equations~$R_{rr} = R_{\theta \theta} = R_{\phi
  \phi}= 0$ linearized about flat space we see the following structure
\begin{subequations}
  \label{eqn:aff_null_spher_jord_block}
  \begin{align}
    &
    \p_r \delta g_{uu} - \frac{1}{2\rho} \p_\rho
    \left(
    \p_r \delta g_{\theta \theta}
    + \sin^{-2} \theta \p_r \delta g_{\phi \phi}
    \right)
    = 0
    \,,
    \label{eqn:aff_null_spher_jord_block_1}
    \\
    &
    \p_r \left(
    \p_r \delta g_{\theta \theta}
    + \sin^{-2} \theta \p_r \delta g_{\phi \phi}
    \right)
    = 0
    \,.
    \label{eqn:aff_null_spher_jord_block_2}
  \end{align}
\end{subequations}
The
variable~$ \left( \p_r \delta g_{\theta \theta} + \sin^{-2} \theta
  \p_r \delta g_{\phi \phi} \right)$
in~\eqref{eqn:aff_null_spher_jord_block_1} prevents~$\delta g_{uu}$
from satisfying just an advection equation along~$\p_r$ and so
provides a nontrivial Jordan block. The combination
of~$\delta g_{\theta \theta}$ and~$\delta g_{\phi \phi}$ in the former
hints that a different choice of variables may be more
appropriate. This combination of variables furthermore appears in the
trivial equation~$R_{ur}=0$ when linearized about flat space, and so
it may be optimal to group them together. We thus next consider the
equations as resulting from the metric ansatz
\begin{align*}
  ds^2 = g_{uu} du^2 -2dudr + \hat{R}(u,r)^2
  \left( d\theta^2 + \sin^2 \theta d\phi^2 \right)
  \,,
\end{align*}
where~$\hat{R}$ is the radius of the two-sphere. This form of the
metric ansatz is used in the spherically symmetric case
of~\cite{Win13}, employed by~\cite{CreOliWin19} in the study of
gravitational collapse of a massless scalar field, as well as
in~\cite{vdWBis12} for cosmological considerations using past null
cones. Upon linearization about flat space the characteristic PDE
system takes the form
\begin{subequations}
  \label{eqn:char_aff_null_spher}
  \begin{align}
  &
    \p_r^2 \delta \hat{R} = 0
  \,,
  \label{eqn:char_aff_null_spher_a}
  \\
  &
    2 r \p_u \p_r \delta \hat{R} + 2 \p_u \delta \hat{R}
    \nonumber
  \\
  & \quad
    -2 \p_r \delta \hat{R} + r \p_r \delta g_{uu} + \delta g_{uu}
    = 0
    \,,
    \label{eqn:char_aff_null_spher_b}
  \\
  &
    4 \p_u \p_r \delta \hat{R} + r \p_r^2 \delta g_{uu} + 2 \p_r \delta g_{uu}
    =0
    \,.
    \label{eqn:char_aff_null_spher_c}
\end{align}
\end{subequations}
Equations~\eqref{eqn:char_aff_null_spher_a}
and~\eqref{eqn:char_aff_null_spher_b} correspond to the main
equations~$R_{rr} = 0$ and~$R_{\theta \theta} =0$, respectively, and
Eq.~\eqref{eqn:char_aff_null_spher_c} to the trivial
one~$R_{ur}=0$. The main equation~$R_{\phi \phi}$ is dropped since it
is proportional to~$R_{\theta \theta}$ and the two-sphere is
parametrized only by its radius.

Comparing once more with the ADM form of the problem, including in the
system the trivial equation~\eqref{eqn:char_aff_null_spher_c}
corresponds to including in the analysis the linearized ADM equation
for~$\delta K_{\rho \rho}$.  This is an essential component in
identifying the pure gauge sub-block along the radial direction. To
achieve this we first make the following identification using
Eq.~\eqref{eqn:gen_BS_vars_to_ADM}:
\begin{align*}
  g_{uu} = \alpha^{-2} - 2\,,
\end{align*}
which after linearization about flat space yields
\begin{align}
  \delta g_{uu} = -2 \delta \alpha = \delta \gamma_{\rho \rho}
  \,,\label{eqn:dguu_gaugefixing}
\end{align}
where the gauge
condition~$\delta \alpha = - \delta \gamma_{\rho \rho}/2$ is used. We
consider now a first order reduction with
\begin{align*}
  (\p_r \delta \hat{R})
  \,,
  \quad
  (\p_u \delta \hat{R})
  \,,
  \quad
  ( \p_r \delta g_{uu})
\end{align*}
promoted to independent variables where,
by~\eqref{eqn:dguu_gaugefixing}, the latter is equivalent
to~$(\p_r \delta \gamma_{\rho \rho})$ being treated as a reduction
variable. This first order reduction provides a diagonalizable radial
principal part for~\eqref{eqn:char_aff_null_spher} with advection
equations along~$\p_r$ for all variables--original and auxiliary--and
corresponds to the pure gauge
subsystem~\eqref{eqn:aff_null_pg_2nd_order}. More precisely, the
relation between the ADM gauge variables and the characteristic
variables is
\begin{subequations}
  \label{eqn:char_vars_to_gauge_ADM_spher}
  \begin{align}
    \tfrac{1}{2} \p_\rho \delta \gamma_{\rho \rho}
    &
      = \tfrac{1}{2}
      ( \p_r \delta g_{uu} - \p_u \delta g_{uu})
      \,,
      \label{eqn:char_vars_to_gauge_ADM_spher_a}
    \\
    - \delta K_{\rho \rho}
    &
      = \p_r \delta g_{uu}
      - \tfrac{1}{2} \p_u \delta g_{uu}
      \,.
      \label{eqn:char_vars_to_gauge_ADM_spher_b}
  \end{align}
\end{subequations}
Since all characteristic variables satisfy advection equations
along~$\p_r$, combining~\eqref{eqn:char_vars_to_gauge_ADM_spher}
with~\eqref{eqn:gauge_ADM_vars_spher_null_dev_a}
and~\eqref{eqn:gauge_ADM_vars_spher_null_dev_b} one
recovers~\eqref{eqn:pure_gauge_sub-block_2nd_order_spher}.

If~$(\p_u \delta g_{uu})$ is also taken as an auxiliary variable, then
the first order reduction is of the standard type, since
\begin{align*}
  \p_\rho \delta g_{uu} = \p_r \delta g_{uu} - \p_u \delta g_{uu}
  \,.
\end{align*}
The equation of motion for~$(\p_u \delta g_{uu})$ can be obtained from
\begin{align*}
  \p_r ( \p_u \delta g_{uu})  = \p_u (\p_r \delta g_{uu})
  \,.
\end{align*}
This first order reduction of~\eqref{eqn:char_aff_null_spher}
possesses the following nontrivial Jordan block
\begin{align*}
  (\p_t+\p_\rho) (\p_u \delta g_{uu} ) + \p_\rho (\p_r \delta g_{uu}) = 0
  \,,
  \\
  (\p_t + \p_\rho) ( \p_r \delta g_{uu} ) = 0
  \,,
\end{align*}
and a linear combination yields
\begin{align*}
  (\p_t + \p_\rho) \left[
  ( \p_r \delta g_{uu}) - (\p_u \delta g_{uu})
  \right]
  = \p_\rho ( \p_r \delta g_{uu} )
\,.
\end{align*}
Via the identification~\eqref{eqn:char_vars_to_gauge_ADM_spher} the
latter matches~\eqref{eqn:ADM_lin_mink_aff_null_spher_3}, modulo an
overall factor of~$1/2$. Hence, the Jordan block of the characteristic
PDE with this characteristic standard first order reduction coincides
precisely with the pure gauge principal part~\eqref{eqn:aff_null_pg_a}
and~\eqref{eqn:aff_null_pg_b}. This is merely the characteristic
version of the standard first order reduction in the Cauchy frame. The
alternative choice, where, instead of introducing
both~$(\p_u \delta g_{uu})$ and~$(\p_r \delta g_{uu})$ as auxiliary
variables, only the latter is introduced, renders the characteristic
PDE system in spherical symmetry strongly hyperbolic. Consequently,
the initial value problem of this system is not well-posed in a norm
where both~$(\p_t \delta g_{uu})^2$ and~$(\p_\rho \delta g_{uu})^2$
are included in the integrand, but in one that involves
only~$ (\p_r \delta g_{uu})^2$. Based on this norm, one can study
well-posedness of the CIBVP of the system by seeking energy estimates,
similar to the analysis of~\cite{GiaHilZil20}. See also~\cite{Bal97}
for energy estimates of the wave and Maxwell equations in a
single-null characteristic setup.

\subsection{Pure gauge sub-block: Angular direction~$\theta$}
\label{Subsec:aff_null:pg_sub-block:theta_dir}

We next expand the previous analysis to a setup without symmetry,
focusing purely on the angular direction~$\theta$. The pure gauge
structure is identified in both the ADM and characteristic setups. In
contrast, however, to the radial direction there is no combination of
variables that allows us to avoid the nontrivial Jordan block of the
pure gauge. We also discuss which choice of variables is most
convenient for the analysis.

\subsubsection{ADM setup}

The partition in to gauge, constraint, and physical variables along
the~$\theta$ direction is still achieved using
Eqs.~\eqref{eqn:gauge_vars},~\eqref{eqn:constr_vars},
and~\eqref{eqn:phys_projector}, respectively. The gauge variables are
\begin{equation}
      \begin{aligned}
  [\p_\theta^2 \Theta]
  &= - \delta K_{\theta \theta}
  \,, \quad \quad \; \;
  [\p_\theta^2 \psi^\rho] = \p_\theta \delta \gamma_{\rho \theta}
  \,, 
  \\
  [\p_\theta^2 \psi^\theta]
  & = \frac{1}{2 \rho^2} \p_\theta \delta \gamma_{\theta \theta}
    \,, \quad
  [\p_\theta^2 \psi^\phi] = \frac{1}{\rho^2 \sin^2 \theta}
    \p_\theta \delta \gamma_{\theta \phi}
      \,.
    \end{aligned}
    \label{eqn:gauge_vars_theta}
  \end{equation}
The constraint variables are
\begin{align}
  &
    [H]
    = -\p_\theta \delta \gamma_{\rho \rho}
    -  \frac{1}{ \rho^2 \sin^2 \theta} \p_\theta \delta \gamma_{\phi \phi}
    \,, \quad
    [M_\rho] = \delta K_{\rho \theta}
    \,,
    \label{eqn:constr_vars_theta}
  \\
  & \! \!
    [M_\theta]
    = - \delta K_{\rho \rho} -  \frac{1}{\rho^2 \sin^2 \theta} \delta K_{\phi \phi}
    \,, \qquad
    [M_\phi] = \delta K_{\theta \phi}
    \,.
    \nonumber
\end{align}
The physical variables are obtained with the action
of~$^P \! \! \!  \perp$ on~$\delta \gamma_{ij}$ and~$\delta
K_{ij}$. As seen from the physical
subsystem~\eqref{eqn:physical_subsys}, the latter is essentially a
time derivative of the former. We work with the physical variables
\begin{subequations}
  \begin{align}
    [h_+]
    &\equiv \frac{1}{2} \delta \gamma_{\rho \rho}
    - \frac{1}{2 \rho^2 \sin^2 \theta} \delta \gamma_{\phi \phi}
    \,, \quad
    [h_\times] \equiv \delta \gamma_{\rho \phi}
    \,,
          \label{eqn:physical_vars_theta_h}
    \\
    [\dot{h}_+] &\equiv
    \frac{1}{ \rho^2 \sin^2 \theta} \delta K_{\phi \phi}
    - \delta K_{\rho \rho}\,, \quad
    [\dot{h}_\times] \equiv -2 \delta K_{\rho \phi}
    \,,
    \label{eqn:physical_vars_theta_hdot}
  \end{align}
  \label{eqn:physical_vars_theta}%
\end{subequations}
which correspond to the two polarizations of the gravitational waves
in GR. In Eq.~\eqref{eqn:physical_vars_theta_hdot} we have multiplied
with an overall factor of~$-2$ for the definitions to be compatible
with the physical subsystem~\eqref{eqn:physical_subsys}
when~$[\dot{h}_+] =\p_t h_+$, and similarly for~$[h_\times]$. As
expected for a gravitational wave that travels along the~$\theta$
direction, the physical variables involve only spatial metric
components that are transverse to this direction. The principal symbol
in the form~\eqref{eqn:second_order_principal_symbol_eom_form} in
the~$\theta$ direction for the linearized ADM formulation is
\begin{subequations}
  \begin{align}
    \p_t \delta \gamma_{\rho \rho}
    &\simeq - 2 \delta K_{\rho \rho}
      \,,
      \label{eqn:delta_gamma_ADM_theta_princ_a}
    \\
    \p_t \delta \gamma_{\rho \theta}
    &\simeq - 2 \delta K_{\rho \theta} -  \p_\theta \delta \gamma_{\rho \rho}
      \,,
      \label{eqn:delta_gamma_ADM_theta_princ_b}
    \\
    \p_t \delta \gamma_{\rho \phi}
    &\simeq - 2 \delta K_{\rho \phi}
      \,,
      \label{eqn:delta_gamma_ADM_theta_princ_c}
    \\
    \p_t \delta \gamma_{\theta \theta}
    &\simeq - 2 \delta K_{\theta \theta} - 2 \p_\theta \delta \gamma_{\rho \theta}
      \,,
      \label{eqn:delta_gamma_ADM_theta_princ_d}
    \\
    \p_t \delta \gamma_{\theta \phi}
    &\simeq - 2 \delta K_{\theta \phi} - \p_\theta \delta \gamma_{\rho \phi}
      \,,
         \label{eqn:delta_gamma_ADM_theta_princ_e}  
    \\
    \p_t \delta \gamma_{\phi \phi}
    &\simeq - 2 \delta K_{\phi \phi}
      \,,
    \label{eqn:delta_gamma_ADM_theta_princ_f}
  \end{align}
  \label{eqn:delta_gamma_ADM_theta_princ}
\end{subequations}
and
\begin{subequations}
  \label{eqn:delta_K_ADM_theta_princ}
  \begin{align}
    \p_t \delta K_{\rho \rho}
    & \simeq - \frac{1}{2 \rho^2} \p_\theta^2 \delta \gamma_{\rho \rho}
      \,,
      \label{eqn:delta_K_ADM_theta_princ_a}
    \\
    \p_t \delta K_{\rho \theta}
    & \simeq 0
      \,,
      \label{eqn:delta_K_ADM_theta_princ_b}
    \\
    \p_t \delta K_{\rho \phi}
    & \simeq - \frac{1}{2 \rho^2} \p_\theta^2 \delta \gamma_{\rho \phi}
      \,,
      \label{eqn:delta_K_ADM_theta_princ_c}
    \\
    \p_t \delta K_{\theta \theta}
    & \simeq - \frac{1}{2 \rho^2 \sin^2 \theta} \p_\theta^2 \delta \gamma_{\phi \phi}
      \,,
      \label{eqn:delta_K_ADM_theta_princ_d}
    \\
    \p_t \delta K_{\theta \phi}
    & \simeq 0
      \,,
      \label{eqn:delta_K_ADM_theta_princ_e}
    \\
    \p_t \delta K_{\phi \phi}
    & \simeq - \frac{1}{2\rho^2} \p_\theta^2 \delta \gamma_{\phi \phi}
      \,.
      \label{eqn:delta_K_ADM_theta_princ_f}
  \end{align}
\end{subequations}
For a standard first order reduction the pure gauge principal
structure along the~$\theta$ direction is inherited by
\begin{subequations}
  \begin{align}
    \p_t (\tfrac{1}{2\rho^2} \p_\theta \delta \gamma_{\theta \theta})
    & \simeq - \rho^{-2} \p_\theta
      \left( \p_\theta \delta \gamma_{\rho \theta}
      + \delta K_{\theta \theta} \right)
      \,,
      \label{eqn:theta_pg_sub-block_jord_a}
    \\
    \p_t \left( \p_\theta \delta \gamma_{\rho \theta} + \delta K_{\theta \theta} \right)
    & \simeq
      - \p_\theta^2 \delta \gamma_{\rho \rho}
      - \tfrac{1}{2 \rho^2 \sin^2 \theta} \p_\theta^2 \delta \gamma_{\phi \phi}
      \nonumber
    \\
    & \quad 
      - 2 \p_\theta \delta K_{\rho \theta}
      \,.
      \label{eqn:theta_pg_sub-block_jord_b}
  \end{align}
  \label{eqn:theta_pg_sub-block_jord}
\end{subequations}
After using
Eqs.\eqref{eqn:gauge_vars_theta},~\eqref{eqn:constr_vars_theta},
and~\eqref{eqn:physical_vars_theta} the
system~\eqref{eqn:theta_pg_sub-block_jord} yields
\begin{equation}
  \begin{aligned}
    & \p_t [\p_\theta^2 \psi^\theta]
    + \rho^{-2}  \p_\theta [\p_\theta^2 \left( \psi^\rho -\Theta \right)]
    \simeq  0
    \,,
    \\
    & \p_t [\p_\theta^2  \left( \psi^\rho - \Theta \right)]
    \simeq \tfrac{3}{4} \p_\theta [H]
    - 2 \p_\theta [M_\theta]
    - \tfrac{1}{2} \p_\theta^2[h_+]
    \,,
  \end{aligned}
  \label{eqn:ADM_pg_subsystem_theta}
\end{equation}
so that, comparing with~\eqref{eqn:aff_null_pg}, the pure gauge
structure of~$\mathbf{P}_G$ is manifest within the full principal
symbol, as too is the coupling among gauge, constraint, and physical
variables encoded in~$\mathbf{P}_{GC}$ and~$\mathbf{P}_{GP}$. Here we
have worked with the plain ADM evolution equations. Working with the
ADM equivalent discussed in Sec.~\ref{Section:BS_to_ADM} changes only
the coupling to the constraints. To obtain this result the necessary
conditions were
\begin{enumerate}
\item Introduction of the quantities
  $(\p_\theta \delta \gamma_{\theta \theta})$
  and~$(\p_\theta \delta \gamma_{\rho \theta})$ as auxiliary
  variables.
\item Inclusion of the equation of motion for~$\delta K_{\theta
  \theta}$ in the analyzed system.
\end{enumerate}
Interestingly, the affine null gauge provides an explicit example
where the sub-block~$\mathbf{P}_{GP}$ of the full principal
symbol~$\mathbf{P}^s$ is nonvanishing, so there is nontrivial coupling
between gauge and physical variables in the principal symbol.

\subsubsection{Characteristic setup}

We repeat now the previous analysis directly in the characteristic
coordinates and variables to demonstrate how the pure gauge structure
is inherited in~$\mathbf{P}^\theta$ for the characteristic setup. The
ADM analysis is again used as guidance in this.  More specifically,
from the equivalent ADM system~\eqref{eqn:equiv_ADM_to_main_sys} we
know that the characteristic system involves the equation of motion
for~$\delta K_{\theta \theta}$, which is one of the two necessary
conditions in order to recover the structure we are looking for. We
parametrize the metric functions simply by
~$g_{uu},\,g_{u\theta},\,g_{u\phi},\,g_{\theta\theta},\,g_{\phi
  \theta},\, g_{\phi \phi}$. For the present calculations this choice,
as opposed to that of~\cite{Win13}, is preferred due to its cleaner
connection to the ADM variables and allows us to uncover the pure
gauge structure more easily.

With this parametrization the PDE system consisting of the main
equations~\eqref{eqn:main_BS_sys} does not involve terms of the
form~$\p_\theta^2 \delta g_{u \theta}$
and~$\p_\theta^2 \delta g_{\theta \theta}$, which in the ADM language
correspond to~$\p_\theta^2 \delta \gamma_{\rho \theta}$
and~$\p_\theta^2 \delta \gamma_{\theta \theta}$. A minimal first order
reduction of the characteristic system, the details of which can be
found in the ancillary files, exhibits the following Jordan block in
the~$\theta$ direction:
\begin{align*}
  &
    \p_t \delta g_{uu}
    + \frac{1}{2 \rho \sin^2 \theta} \p_t (\p_r \delta g_{\theta \theta})
  \\
  & \qquad \quad
    - \frac{1}{\rho^2} \p_\theta \delta g_{u \theta}
    + \frac{\cot \, \theta}{2 \rho^3} \p_\theta \delta g_{\theta \theta}
    \simeq 0
    \,,
  \\
  & \frac{1}{\rho^2} \p_t \delta g_{u \theta}
    - \frac{\cot \, \theta}{2 \rho^3} \p_t \delta g_{\theta \theta}
    \simeq 0 \,.
\end{align*}
This reduction is minimal in the sense that the minimum number of
auxiliary variables needed to form a complete first order system were
introduced. The above structure motivates the introduction
of~$(\p_\theta \delta g_{u \theta})$
and~$(\p_\theta \delta g_{\theta \theta})$ as auxiliary variables in
addition to the minimum, since they form the nontrivial Jordan
block. But, as we saw earlier, this is the other necessary condition
to recover the pure gauge structure in the full system. Thus in the
new first order reduction the~$2 \times 2$ Jordan block along
the~$\theta$ direction persists, namely
\begin{subequations}
  \begin{align}
    &
      \p_t ( \p_\theta \delta g_{\theta \theta})
    - \rho^2 \p_t (\p_r \delta g_{u \theta})
      \label{eqn:theta_pg_sub-block_char_aff_null_a}
    \\
    & \qquad \quad \; \; \,
      - \p_\theta (\p_r \delta g_{\theta \theta})
    - \frac{1}{\sin^2\, \theta} \p_\theta (\p_r \delta g_{\phi \phi})
      \simeq 0
      \,,
      \nonumber
    \\
    &
      \p_t (\p_r \delta g_{\theta \theta})
      + \frac{1}{\sin^2 \, \theta} \p_t (\p_r \delta g_{\phi \phi})
      \simeq 0
      \,.
      \label{eqn:theta_pg_sub-block_char_aff_null_b}
  \end{align}
  \label{eqn:theta_pg_sub-block_char_aff_null}%
\end{subequations}
The latter is indeed the pure gauge sub-block expected from the ADM
analysis. To realize this explicitly we first express the
characteristic auxiliary variables in terms of the ADM ones:
  \begin{align*}
    \p_\theta \delta g_{\theta \theta}
    & =
      \p_\theta \delta \gamma_{\theta \theta}
      \,,
    \\
    \p_r \delta g_{\theta \theta}
    & =
      (\p_t + \p_\rho) \delta \gamma_{\theta \theta}
      \simeq -2 \delta K_{\theta \theta} - 2 \p_\theta \delta \gamma_{\rho \theta}
      \,,
    \\
    \p_r \delta g_{u \theta}
    & = (\p_t + \p_\rho) \delta \gamma_{\rho \theta}
      \simeq -2 \delta K_{\rho \theta} - \p_\theta \delta \gamma_{\rho \rho}
      \,,
    \\
    \p_r \delta g_{\phi \phi}
    & = (\p_t + \p_\rho) \delta \gamma_{\phi \phi}
      \simeq - 2 \delta K_{\phi \phi}
      \,,
  \end{align*}
  where we have dropped derivatives transverse to~$\p_{\theta}$. Then,
  Eq.~\eqref{eqn:theta_pg_sub-block_char_aff_null} reads
\begin{align*}
  &
    \p_t \p_\theta \delta \gamma_{\theta \theta}
    + 2 \rho^2 \p_t \delta K_{\rho \theta}
    + \rho^2 \p_\theta \p_t \delta \gamma_{\rho \rho}
  \\
  & \quad
    + 2 \p_\theta \delta K_{\theta \theta}
    + 2 \p_\theta^2 \delta \gamma_{\rho \theta}
    + \frac{2}{\sin^2 \, \theta} \p_\theta \delta K_{\phi \phi}
    \simeq 0
    \,,
    \nonumber
  \\
  &
    \p_t \delta K_{\theta \theta}
    + \p_t \p_\theta \delta \gamma_{\rho \theta}
    + \frac{1}{\sin^2 \, \theta} \p_t \delta K_{\phi \phi}
    \simeq 0
\end{align*}
which after replacing~$\p_t \delta \gamma_{\rho
  \rho}$,~$\p_t \delta K_{\rho \theta}$,~$\p_t \delta K_{\phi \phi}$
with the right-hand side of~\eqref{eqn:delta_gamma_ADM_theta_princ_a},
\eqref{eqn:delta_K_ADM_theta_princ_b},
\eqref{eqn:delta_K_ADM_theta_princ_f}, respectively, yields
\begin{subequations}
  \begin{align}
    &
      \p_t \left( \frac{1}{2 \rho^2} \p_\theta \delta \gamma_{\theta \theta} \right)
      + \rho^{-2}
      \p_\theta ( \delta K_{\theta \theta} + \p_\theta \delta \gamma_{\rho \theta})
      \simeq
      \label{eqn:theta_pg_sub-block_char_aff_null_2_a}
    \\
    & \qquad \qquad  \qquad 
      \p_\theta \delta K_{\rho \rho}
      - \frac{1}{\rho^2 \sin^2 \, \theta} \p_\theta \delta K_{\phi \phi}
      \,,
      \nonumber
    \\
    &
      \p_t ( \delta K_{\theta \theta} + \p_\theta \delta \gamma_{\rho \theta})
      \simeq
      \frac{1}{2 \rho^2 \sin^2 \, \theta} \p_\theta^2 \delta \gamma_{\phi \phi}
      \,,
      \label{eqn:theta_pg_sub-block_char_aff_null_2_b}
  \end{align}
  \label{eqn:theta_pg_sub-block_char_aff_null_2}%
\end{subequations}
where in~\eqref{eqn:theta_pg_sub-block_char_aff_null_2_a} we have
multiplied overall with a factor of~$1/2 \rho^2$. The right-hand side
of~\eqref{eqn:theta_pg_sub-block_char_aff_null_2} involves only
constraint and physical variables along the~$\theta$ direction, while
the left-hand side shows the coupling only between gauge
variables. Using the
relations~\eqref{eqn:gauge_vars_theta},~\eqref{eqn:constr_vars_theta},
and~\eqref{eqn:physical_vars_theta} the
system~\eqref{eqn:theta_pg_sub-block_char_aff_null_2} reads
\begin{subequations}
  \label{eqn:theta_pg_sub-block_char_aff_null_3}%
  \begin{align}
    &
    \p_t [ \p_\theta^2  \psi^\theta]
    + \rho^{-2}
    \p_\theta [ \p_\theta^2 ( \psi^\rho - \Theta)]
    \simeq
    - \p_\theta [\dot{h}_+]
    \label{eqn:theta_pg_sub-block_char_aff_null_3_a}
    \,,
    \\
    &
    \p_t [ \p_\theta^2  ( \psi^\rho - \Theta)]
    \simeq
    - \frac{1}{4} \p_\theta [H]
    + \frac{1}{2} \p_\theta^2 [h_+]
    \,,
    \label{eqn:theta_pg_sub-block_char_aff_null_3_b}
  \end{align}
\end{subequations}
which again inherits the structure of the pure gauge subsystem, namely
the Jordan block~\eqref{eqn:aff_null_pg_a}
and~\eqref{eqn:aff_null_pg_c}, and provides nontrivial coupling of
gauge to constraint and physical variables. Hence, the nontrivial
Jordan block of~$\mathbf{P}^\theta$ in the characteristic affine null
system corresponds precisely to the non-trivial Jordan block of the
pure gauge subsystem~\eqref{eqn:aff_null_pg} along the same
direction. Comparing the
form~\eqref{eqn:theta_pg_sub-block_char_aff_null_3} to the
form~\eqref{eqn:ADM_pg_subsystem_theta} in the ADM setup, the only
difference is in the coupling of gauge variables to constraint and
physical ones.

A different choice of variables that makes use of
definition~\eqref{eqn:gauge_cond_det} is common in affine null
formulations. Such a choice can, however, make less clear the
distinction among gauge, constraint, and physical variables.  In the
ancillary files we include analyses where we explore such
parametrizations. Crucially, the principal symbol of the
characteristic system is still nondiagonalizable along~$\theta,\phi$,
but the choice of variables is inconvenient in identifying the
different sub-blocks.

\section{More Bondi-like gauges}
\label{Section:more_Bondi-like_gauges}

In this section we repeat the previous analysis for the Bondi-Sachs
gauge proper~\cite{BonBurMet62,Sac62} in the ADM setup. This specific
system is already shown to be WH~\cite{GiaHilZil20}. Again we identify
the nontrivial Jordan block of the full system to that of the pure
gauge subsystem. Additionally, we present the pure gauge subsystem of
the double null gauge and show that it is also only WH. We argue that
the full system in the double null as well as other Bondi-like gauges
is necessarily WH when up to second order metric derivatives are
considered.

\subsection{Bondi-Sachs gauge proper}
\label{Subsection:BS_proper_gauge}

In the outgoing Bondi-Sachs proper gauge the coordinate light speed
conditions~$c_+^\rho=1,\,c_+^A=0$ are imposed--as in all outgoing
Bondi-like gauges--and lead to
\begin{align*}
  \alpha L^{-1} - \beta^\rho =1
  \,,
  \qquad
  \beta^A = - b^A \alpha L^{-1}
  \,,
\end{align*}
in terms of lapse and shift. The gauge is closed by setting
\begin{align}
  \rho = \hat{R}  \,.
  \label{eqn:Bondi_Radius}
\end{align} 
In this form the gauge fixing is not so easily expressed in an ADM
setup, since we do not have a complete specification of the lapse and
shift. We can, however, achieve this by combining the ADM
equations~\eqref{eqn:ADM_lin_mink}, the~$2+1$
split~\eqref{eqn:2+1_gamma_split} of the spatial metric~$\gamma_{ij}$
and the determinant condition~\eqref{eqn:Bondi_Radius}. We basically
want to specify a~$\beta^\rho$ for which the determinant
condition~\eqref{eqn:Bondi_Radius} is satisfied at later
times. Starting from the standard ADM equations on the two-sphere we
get
\begin{align}
  \mathcal{L}_t q_{AB}
    & = - 2 \alpha \, \prescript{(q)\!\!}{}\perp K_{AB}
      \label{eqn:angular_ADM_eom}
  \\
    & \quad \,
      + \mathcal{L}_{[\beta^\rho \p_\rho]\,} q_{AB}
      - \mathcal{L}_{[(1+\beta^\rho) b]\,}
      q_{AB}
      \,,
      \nonumber
\end{align} 
where~$\prescript{(q)\!\!}{}\perp$ denotes the projection with respect
to~$q_{AB}$ on every open index and~$b^a$ denotes the slip vector. The
general relation between the derivative of a matrix and the derivative
of its determinant applied to~$q_{AB}$ yields
\begin{align*}
  q^{ab} \mathcal{L}_t q_{ab} = q^{ab} \p_t q_{ab} = \p_t \ln(q)
  \,,
\end{align*}
where~$q \equiv \det(q)$. Imposing the determinant
condition~\eqref{eqn:Bondi_Radius} the latter
yields~$q^{ab} \mathcal{L}_t q_{ab} = 0$. Then,
Eq.~\eqref{eqn:angular_ADM_eom} after tracing with~$q^{AB}$ returns
\begin{align*}
  0
  & =
    -2 \alpha K_{qq}
    + \beta^\rho \left[
    \p_\rho \ln(q) - 2 \slashed{D}_A b^A  
    \right] - 2 \slashed{D}_A b^A
    \,,
\end{align*}
where~$\slashed{D}_A$ is the covariant derivative compatible
with~$q_{AB}$. Using~$c_+^\rho=1=-\beta^\rho+\alpha/L$ we finally
obtain~$\beta^\rho = \rho \, X /(4 - \rho \, X)$ with
\begin{align*}
  X = 2 L K_{qq} + 2 \slashed{D}_a b^a
\end{align*}
and~$\p_\rho \ln(q) = 4/\rho$. In terms of the lapse and shift the
Bondi-Sachs proper gauge can thus be imposed by
\begin{equation}
  \begin{aligned}
    \alpha
    &= L (1 + \beta^\rho)
    \,, \quad
    \beta^\rho
    &= \frac{X \, \rho/4}{1 - X \, \rho/4}
    \\
    \beta^\theta
    &= - b^\theta \alpha L^{-1}
    \,, \quad
    \beta^\phi
    &= - b^\phi \alpha L^{-1}
    \,,
  \end{aligned}
  \label{eqn:BS_gauge_2}
\end{equation}
which is a mixed algebraic-differential gauge.

\subsubsection{Pure gauge subsystem}
\label{subsubsection:BS_gauge:pg_subsys}

To proceed with our analysis we first need to obtain the pure gauge
subsystem~\eqref{eqn:PG_subsys_lin_mink} for the Bondi-Sachs gauge.
We continue in the linear constant coefficient approximation. Under
this assumption the Bondi-Sachs proper gauge~\eqref{eqn:BS_gauge_2}
reads
\begin{equation}
\label{eqn:BS_gauge_lin_mink}
\begin{aligned}
  \delta \alpha
  &= \delta \beta^\rho+ \frac{1}{2} \delta \gamma_{\rho \rho}
    \,,
  \\
  \delta \beta^\rho
  & = \frac{\delta K_{\theta \theta}}{2\rho}
    + \frac{\delta K_{\phi \phi}}{2 \rho \sin^2 \theta}
    + \frac{\p_\theta \delta \gamma_{\rho \theta}}{2\rho}
    + \frac{\p_\phi \delta \gamma_{\rho \phi}}{2 \rho \sin^2 \theta} \\
    \qquad &{}+ \frac{\cot \theta \, \delta \gamma_{\rho \theta}}{2 \rho}
    \,,
  \\
  \delta \beta^\theta
  &= - \rho^{-2} \delta \gamma_{\rho \theta}
    \,,
  \\
  \delta \beta^\phi
  &= - (\rho \sin \theta)^{-2} \delta \gamma_{\rho \phi}
    \,.
  \end{aligned}
\end{equation}
Replacing these in Eq.~\eqref{eqn:PG_subsys_lin_mink} and using the
relations~\eqref{eqn:gauge_vars_theta} to translate to the gauge
variables, the pure gauge subsystem of the Bondi-Sachs proper gauge
reads
\begin{equation}
      \label{eqn:BS_PG_sys_2nd_order}
\begin{aligned}
  & \p_t \Theta
+ \frac{1}{2\rho} \p_\theta^2 \Theta
    + \frac{1}{2 \rho \sin^2 \theta} \p_\phi^2 \Theta
    - \frac{1}{2 \rho} \p_\theta^2 \psi^\rho
    - \frac{1}{2 \rho \sin^2 \theta} \p_\phi^2 \psi^\rho \\
  & \quad \; \; \,
    - \frac{\rho}{2} \p_\rho \p_\theta \psi^\theta
    - \frac{\rho}{2} \p_\rho \p_\phi \psi^\phi
    - \p_\rho \psi^\rho
    - \frac{\cot \theta}{2\rho} \p_\theta \psi^\rho \\
  & \quad \; \; \,
    - \frac{\rho \cot \theta}{2} \p_\rho \psi^\theta
    = 0\,,
  \\
  &
    \p_t \psi^\rho
    + \frac{1}{2\rho} \p_\theta^2 \Theta
    + \frac{1}{2 \rho \sin^2 \theta} \p_\phi^2 \Theta
    - \frac{1}{2 \rho} \p_\theta^2 \psi^\rho
    - \frac{1}{2 \rho \sin^2 \theta} \p_\phi^2 \psi^\rho \\
  & \quad \; \; \; \,
    - \frac{\rho}{2} \p_\rho \p_\theta \psi^\theta
    - \frac{\rho}{2} \p_\rho \p_\phi \psi^\phi
    - \p_\rho \Theta
    - \frac{\cot \theta}{2\rho} \p_\theta \psi^\rho \\
  & \quad \; \; \; \,
    - \frac{\rho \cot \theta}{2} \psi_\rho \psi^\theta
    = 0\,, \\
  &
    \p_t \psi^\theta
    + \p_\rho \psi^\theta
    + \rho^{-2} \p_\theta ( \psi^\rho - \Theta) = 0\,, \\
  &\p_t \psi^\phi
    + \p_\rho \psi^\phi
    + (\rho \sin \theta)^{-2} \p_\phi (\psi^\rho -\Theta)
    =0\,.
  \end{aligned}
\end{equation}
To analyze the hyperbolicity of this second order in space system we
consider a first order reduction with variables
\begin{align*}
  &
  \Theta-\psi^\rho
  ,\,
  \p_\theta(\Theta-\psi^\rho)
  ,\,
  \p_\phi(\Theta-\psi^\rho)
  \,,
    \nonumber
  \\
  & \Theta+\psi^\rho
  ,\,
  \psi^\theta
  \,,
  \p_\theta \psi^\theta
  \,,
  \psi^\phi
  \,,
  \p_\phi \psi^\phi
    \,.
\end{align*}
The minimal first order reduction of this system reads
\begin{subequations}
  \begin{align}
    &
      \p_t( \Theta - \psi^\rho) + \p_\rho(\Theta - \psi^\rho)
      =0
      \,,
      \label{eqn:BS_PG_sys_1st_order_a}
    \\
    &
      \p_t [\p_\theta( \Theta - \psi^\rho)] + \p_\rho [\p_\theta(\Theta - \psi^\rho)]
      =0
      \,,
      \label{eqn:BS_PG_sys_1st_order_b}
    \\
    &
      \p_t [\p_\phi( \Theta - \psi^\rho)] + \p_\rho [\p_\phi(\Theta - \psi^\rho)]
      =0
      \,,
      \label{eqn:BS_PG_sys_1st_order_c}
    \\
    &
      \p_t( \Theta + \psi^\rho) - \p_\rho( \Theta+\psi^\rho)
      -\frac{\cot \theta}{2 \rho} \p_\theta(\Theta+\psi^\rho)
      \nonumber
    \\
    & \quad
      +\rho^{-1} \p_\theta [ \p_\theta(\Theta-\psi^\rho)]
      + \rho^{-1} \sin^{-2}\theta \p_\phi [\p_\phi(\Theta - \psi^\rho)]
      \nonumber
    \\
    & \quad
      +\frac{\cot \theta}{2 \rho} \p_\theta(\Theta-\psi^\rho)
      + \rho \cot \theta \p_\rho \psi^\theta
      \nonumber
    \\
    & \quad
      - \rho \p_\rho (\p_\theta \psi^\theta)
      - \rho \p_\rho (\p_\phi \psi^\phi)
      = 0
      \,,
      \label{eqn:BS_PG_sys_1st_order_d}
    \\
    &
      \p_t \psi^\theta + \p_\rho \psi^\theta
      -\rho^{-2}[\p_\theta(\Theta-\psi^\rho)]
      = 0
      \,,
      \label{eqn:BS_PG_sys_1st_order_e}
    \\
    &
      \p_t (\p_\theta\psi^\theta) + \p_\rho (\p_\theta \psi^\theta)
      - \rho^{-2}\p_\theta [\p_\theta ( \Theta - \psi^\rho)]
      = 0
      \,,
      \label{eqn:BS_PG_sys_1st_order_f}
    \\
    &
      \p_t \psi^\phi + \p_\rho \psi^\phi
      -\rho^{-2}\sin^{-2}\theta [\p_\phi(\Theta-\psi^\phi)]
      = 0
      \,,
      \label{eqn:BS_PG_sys_1st_order_g}
    \\
    &
      \p_t (\p_\phi \psi^\phi) + \p_\rho (\p_\phi \psi^\phi)
    \nonumber \\
    & \quad
      - (\rho \sin \theta)^{-2} \p_\phi [ \p_\phi ( \Theta - \psi^\rho)]
      = 0
      \,.
      \label{eqn:BS_PG_sys_1st_order_h}
  \end{align}
  \label{eqn:BS_PG_sys_1st_order}%
\end{subequations}
All principal matrices of this system possess real eigenvalues, but
the angular principal matrices are nondiagonalizable. The nontrivial
Jordan block along the~$\theta$ direction is given by (see ancillary
files)
\begin{align*}
  &\p_t[\p_\theta(\Theta-\psi^\rho)]
  \simeq 0
  \,,
  \\
  &\p_t (\p_\theta \psi^\theta)
  - \rho^{-2} \p_\theta[\p_\theta(\Theta-\psi^\rho)]
  \simeq 0
  \,,
\end{align*}
and similarly along~$\phi$ by
\begin{align*}
  &\p_t (\p_\phi \psi^\phi)
    - \rho^{-2}\sin^{-2}\theta \p_\phi[\p_\phi(\Theta-\psi^\rho)]
    \simeq 0
    \,,
  \\
  &\p_t[\p_\phi(\Theta-\psi^\rho)]
    \simeq 0
    \,.
\end{align*}
As in the PDE analysis of~\cite{GiaHilZil20} for the axisymmetric
characteristic Bondi-Sachs system, the coupled generalized
characteristic variables obtained here effectively involve second
order angular derivatives. Hence, they cannot be removed with a
different first order reduction of the second order
system~\eqref{eqn:BS_PG_sys_2nd_order}. Thus, the analysis based on
the minimal reduction just performed suffices to show that the pure
gauge subsystem of the Bondi-Sachs proper
gauge~\eqref{eqn:BS_PG_sys_2nd_order} is only WH.

\subsubsection{Pure gauge sub-block: Angular direction~$\theta$}
\label{subsubsection:BS:pg_sub-block:theta_dir}

Similar to Sec.~\ref{Subsec:aff_null:pg_sub-block:theta_dir} we
present the set of evolution equations that inherit the structure of
the pure gauge subsystem in the ADM setup. The necessary conditions to
uncover this structure remain the same. The system that captures the
structure of the pure gauge subsytem along the~$\theta$ direction is
\begin{subequations}
  \begin{align}
    &-\p_t \left(
      \delta K_{\theta \theta} + \p_\theta \delta \gamma_{\rho \theta}
      \right)
      \simeq
      \frac{1}{2} \p_\theta^2 \delta \gamma_{\rho \rho}
      + 2 \p_\theta \delta K_{\rho \theta}
      \label{eqn:ADM_theta_PG_sub-block_a}
    \\
    & \qquad \qquad \qquad
      + \frac{1}{2} \p_\theta^2 \delta \gamma_{\rho \rho}
      + \frac{1}{2\rho^2 \sin^2 \theta} \p_\theta^2 \delta \gamma_{\phi \phi}
      \,, \nonumber
    \\
    &-\p_t \left(
      \delta K_{\theta \theta} - \p_\theta \delta \gamma_{\rho \theta}
      \right)
      \simeq
      \frac{1}{2} \p_\theta^2 \delta \gamma_{\rho \rho}
      - 2 \p_\theta \delta K_{\rho \theta}
      \label{eqn:ADM_theta_PG_sub-block_b}
    \\
    & \qquad
      + \frac{1}{2} \p_\theta^2 \delta \gamma_{\rho \rho}
      + \frac{1}{2\rho^2 \sin^2 \theta} \p_\theta^2 \delta \gamma_{\phi \phi}
      + \p_\theta^2 \delta \beta_\rho
      \,,
      \nonumber
    \\
    &\frac{1}{2\rho^2}  \p_t (\p_\theta \delta \gamma_{\theta \theta})
      \simeq
      - \frac{1}{\rho^2} \p_\theta \delta K_{\theta \theta}
      + \frac{1}{\rho^2} \p_\theta^2 \delta \beta_\theta
      \,,
      \label{eqn:ADM_theta_PG_sub-block_c}
    \\
    &\frac{1}{\rho^2 \sin^2 \theta}  \p_t (\p_\theta \delta \gamma_{\theta \phi})
      \simeq
      \label{eqn:ADM_theta_PG_sub-block_d}
    \\
    & \qquad
      \frac{-2}{\rho^2 \sin^2 \theta} \p_\theta \delta K_{\theta \phi}
      + \frac{1}{\rho^2 \sin^2 \theta} \p_\theta^2  \delta \beta_\phi
      \,,\nonumber
  \end{align}
  \label{eqn:ADM_theta_PG_sub-block}%
\end{subequations}
where spatial derivatives transverse to~$\theta$ are dropped. This
system results from linear combinations of the linearized about flat
space ADM equations and does not include equations outside the main
system~\eqref{eqn:main_BS_sys}. Combining
Eqs.~\eqref{eqn:BS_gauge_lin_mink}, \eqref{eqn:gauge_vars_theta},
\eqref{eqn:constr_vars_theta}, \eqref{eqn:physical_vars_theta},
and~\eqref{eqn:Cons_subsys_lin_mink}, the
system~\eqref{eqn:ADM_theta_PG_sub-block} yields
\begin{subequations}
  \begin{align}
    &
      \p_t [\p_\theta^2 (\Theta - \psi^\rho)]
      \simeq
      -\frac{3}{4} \p_\theta [H]
      + 2 \p_\theta [M_\rho]
      + \frac{1}{2}\p_\theta^2 [h_+]
      \,,
      \label{eqn:ADM_theta_PG_sub-block_2_a}
    \\
    &
      \p_t [\p_\theta^2  (\Theta + \psi^\rho)]
      \simeq
      - \rho^{-1} \p_\theta^2 [ \p_\theta^2(\Theta-\psi^\rho)]
      - \frac{\cot \theta}{\rho} \p_\theta [\p_\theta^2 \psi^\rho]
      \nonumber
    \\
    & \qquad \qquad \qquad \quad \; \,
      - 2 \p_\theta [M_\rho]
      - \frac{3}{4} \p_\theta [H]
      + \frac{1}{2} \p_\theta^2 [h_+]
      \nonumber
    \\
    & \qquad \qquad \qquad \quad \; \,
      - \frac{3}{2} \p_\theta^2 [M_\theta]
      + \frac{1}{2} \p_\theta^2 [\dot{h}_+]
      \,,
      \label{eqn:ADM_theta_PG_sub-block_2_b}
    \\
    &
      \p_t [\p_\theta^2 \psi^\theta]
      \simeq
      \rho^{-2} \p_\theta [\p_\theta^2(\Theta-\psi^\rho)]
      \,,
      \label{eqn:ADM_theta_PG_sub-block_2_c}
    \\
    &
      \p_t [\p_\theta^2 \psi^\phi]
      \simeq
      \frac{-2}{\rho^2 \sin^2 \theta} \p_\theta [M_\theta]
      + \frac{1}{\rho^2 \sin^2 \theta}  \p_\theta^2 [h_\times]
      \,.
      \label{eqn:ADM_theta_PG_sub-block_2_d}
  \end{align}
  \label{eqn:ADM_theta_PG_sub-block_2}%
\end{subequations}
To see how this system inherits the structure of the pure gauge
subsystem~\eqref{eqn:BS_PG_sys_1st_order}, let us neglect all nongauge
variables. Let us furthermore consider adding to the system the
following equations:~$\p_\theta$ of
\eqref{eqn:ADM_theta_PG_sub-block_a},~$\p_\phi$ of
\eqref{eqn:ADM_theta_PG_sub-block_a}, ~$\p_\theta$ of
\eqref{eqn:ADM_theta_PG_sub-block_c}, and~$\p_\phi$ of
\eqref{eqn:ADM_theta_PG_sub-block_d}. As seen from the
form~\eqref{eqn:ADM_theta_PG_sub-block_2} these additional equations
provide the identification to Eq.~\eqref{eqn:BS_PG_sys_1st_order_b},
\eqref{eqn:BS_PG_sys_1st_order_c}, \eqref{eqn:BS_PG_sys_1st_order_f},
and~\eqref{eqn:BS_PG_sys_1st_order_h}, respectively, i.e., the
equations of the auxiliary variables introduced by the minimal first
order reduction. The resulting system is an overall~$\p_\theta^2$
derivative of the first order reduced pure gauge
subsystem~\eqref{eqn:BS_PG_sys_1st_order}. Thus, the hyperbolic
character of the sub-block~$\mathbf{P}_G$ is that of the pure gauge
subsystem, which is WH. Furthermore, from the
form~\eqref{eqn:ADM_theta_PG_sub-block_2} we see another explicit
example of a Bondi-like gauge where~$\mathbf{P}_{GP} \neq
0$. Identification of the pure gauge structure directly in the
characteristic setup is messy with this radial coordinate, so we do
not discuss it in detail.

\subsection{Double-null and more gauges}
\label{Subsection:double_null_gauge}

Another common choice is to use double null coordinates. This was used
in~\cite{Ren90, Chr08, Luk11} to construct initial data on
intersecting ingoing and outgoing null
hypersufaces. Reference~\cite{Ren90} provided the first well-posedness
result to our knowledge for the CIVP in the region near the
intersection, using the harmonic gauge though for the evolution
system, which is symmetric hyperbolic. Reference~\cite{Luk11} improved
this result including in the analysis metric derivatives higher than
second order. A similar approach was used in~\cite{Chr08} as well to
analyze the mathematical conditions for black hole
formation. Norm-type estimates are, of course, central in these
studies, but they are obtained using PDE systems that are not of the
free evolution type and for which the hyperbolic character is not
manifest. If instead one is interested in analyzing a free evolution
system---which is the topic of the current study---then a certain
subset of the systems used in~\cite{Chr08,Luk11} has to be
extracted. There are different choices on how to construct this
subsystem, and in~\cite{HilValZha19} a specific one was shown to
provide a symmetric hyperbolic free evolution scheme in double-null
coordinates. To the best of our knowledge, an evolution scheme with up
to second order metric derivatives using the double null gauge choice
has been used numerically only in spherical symmetry~\cite{Gar95,
  GunBauHil19}.

Working with~$f(\rho)=\rho$ in the coordinate
transformation~\eqref{eqn:coord_transf}, the conditions~$g^{uu}=0$
and~$g^{rr}=0$ yield
\begin{align}
  (\beta^\rho+1)^2 = \alpha^2 \gamma^{\rho \rho}
  \,, \quad
  (\beta^\rho-1)^2 = \alpha^2 \gamma^{\rho \rho}
  \,,
  \label{eqn:double_null_condition_ur}
\end{align}
where the first is the former of the
conditions~\eqref{eqn:gauge_cond_null_gamma_uu} with~$f'=1$. The
conditions~$g^{uA}=0$ are still imposed in the double null gauge,
which provide the latter of
conditions~\eqref{eqn:gauge_cond_null_gamma_uu} with~$f'=1$. From the
coordinate light speed
expressions~\eqref{eqn:radial_coord_light_speed} the
conditions~\eqref{eqn:double_null_condition_ur} yield
\begin{align*}
  c_+^\rho = \pm 1 \,, \quad c_-^\rho = \mp 1
  \,.
\end{align*}
We choose to set~$c_+^\rho = 1$ and~$c_-^\rho = -1$.
Then,~$c_+^\rho + c_-^\rho =0 = -2 \beta^\rho$
implies~$\beta^\rho = 0$, which
from~\eqref{eqn:double_null_condition_ur} leads to~$\alpha =
L$. Replacing these in the second of
conditions~\eqref{eqn:double_null_condition_ur} with~$f'=1$ and
using~\eqref{eqn:slip_vector_relations} provides
$\beta^A = - b^A \alpha L^{-1}$. Then, the whole set of the coordinate
light speeds~\eqref{eqn:radial_coord_light_speed}
and~\eqref{eqn:transverse_coord_light_speed} in the double null gauge
reads
\begin{align*}
  c_+^\rho = 1 \,, \quad
  c_-^\rho = -1 \,,
  \quad
  c_+^A = 0
  \,.
\end{align*}
After linearization about Minkowski, lapse and shift perturbations
read
\begin{align*}
  \delta \alpha
  &= -\frac{1}{2} \delta \gamma_{\rho \rho}
    \,, \quad \quad
    \delta \beta^\theta
    = - \rho^{-2} \delta \gamma_{\rho \theta}
  \\
  \delta \beta^\rho
  &= 0
    \qquad \qquad \quad \; \
    \delta \beta^{\phi} = - \rho^{-2} \sin^2\theta \,
    \delta \gamma_{\rho \phi}
    \,.
    \nonumber
\end{align*}
In terms of~$\Theta$ and~$\psi^i$ the above is similar
to~\eqref{eqn:aff_null_coord_pert} with the only difference that
here~$\delta \beta^\rho = 0$. Then, the pure gauge
subsystem~\eqref{eqn:PG_subsys_lin_mink} for the double null gauge
choice reads
\begin{align*}
  &
    \p_t \Theta - \p_\rho \psi^\rho =  0
    \,,
  \\
  &
    \p_t \psi^\rho - \p_\rho \Theta =  0
    \,,
  \\
  &
    \p_t \psi^\theta + \p_\rho \psi^\theta
    + \rho^{-2} \p_\theta (\psi^\rho - \Theta) =  0
    \,,
  \\
  &
    \p_t \psi^\phi + \p_\rho \psi^\phi
    + (\rho \sin \theta)^{-2} \p_\phi (\psi^\rho - \Theta) =  0
    \,,
\end{align*}
which again possesses nontrivial Jordan blocks along the~$\theta$
and~$\phi$ directions and so is only WH. This was expected since the
difference among the affine null, Bondi-Sachs proper, and double
null cases with respect to the lapse and shift is only in the
specification of the radial coordinate.

This structure in the pure gauge subsystem of the double null gauge
was already discovered in~\cite{Hil15}. We review it here in order to
stress its differences and similarities with other Bondi-like
gauges. We observe that in all three examples that are presented, the
gauge choice~$\beta^A = - b^A \alpha L^{-1}$ renders the pure gauge
subsystem only WH. This choice implies the condition~$c_+^A=0$. Thus
the pure gauge subsystem will also be WH if~$c_-^A=0$ is instead
imposed. In such a case the difference would be a sign change in the
nontrivial Jordan block along the angular directions. Furthermore,
since the specific nature of the angular coordinates (i.e. coordinates
on a two-sphere) is not essential to the WH, we expect that the pure
gauge subsystem would retain this structure if these coordinates
parametrize level sets of a different topology. Our expectation is the
same for higher dimensional spacetimes. In fact, in~\cite{GiaHilZil20}
it was shown that the full characteristic system in the affine null
gauge is WH for a five-dimensional asymptotically AdS spacetime with
planar symmetry. The value of the cosmological constant does not
affect the principal part of the EFEs and so neither their hyperbolic
character.

In summary, we expect that formulations that result from the EFE,
including up to second order metric derivatives will be at best WH if
they are formulated in a Bondi-like gauge. The claim is based on the
following:
\begin{enumerate}
\item The system admits an equivalent ADM setup.
\item The principal symbol~$\mathbf{P}^s$ has the upper triangular
  form~\eqref{eqn:princ_symbol_triang}.
\item The pure gauge sub-block~$\mathbf{P}_G$ inherits the structure
  of the pure gauge subsystem.
\item The pure gauge subsystem is WH.
\end{enumerate}

\section{Numerical experiments}
\label{Section:numerics}

In this section we present convergence tests of the publicly available
characteristic code~\texttt{PITTNULL}~\cite{HanSzi14} which employs
the Bondi-Sachs formalism and is part of the~\texttt{Einstein
  Toolkit}~\cite{einsteintoolkitzenodo}. Although similar tests have
been successfully performed in the past~\cite{ZloGomHus03,
  BabBisSzi09, BabSziWin11, HanSzi14}, the novelty here is that we
examine the convergence of solutions to the full discretized PDE
problem and not just the individual grid functions. The motivation for
this comes from the fact that well-posedness is a property of the full
PDE problem. We examine the practical consequence of the foregoing
results by performing convergence tests in a discretized version of
the~$L^2$ norm. The specific form of that norm plays a key role,
depends on the geometric setup and is inspired by a hyperbolicity
analysis of the PDE system solved. This analysis is similar to that
of~\cite{GiaHilZil20} and can be found in the ancillary files. The
data illustrated in Figs.~\ref{Fig:rescaled_norms_all}
and~\ref{Fig:null_noisy_rescaled_norms} can be found
in~\cite{GiaBisHil21_public}.

\subsection{The setup}

Here we collect the fundamental elements on which
the~\texttt{PITTNULL} code is based. The interested reader can find
more details e.g.\ in~\cite{BisGomLeh97a, HanSzi14}. The Bondi-Sachs
metric ansatz~\cite{BonBurMet62,Sac62} used has the form
\begin{align}
  ds^2
  & =
    - \left( e^{2\beta} \frac{V}{r} - r^2 h_{AB} U^A U^B \right)du^2
    - 2 e^{2\beta} du dr
  \nonumber \\
  & \quad
    - 2 r^2 h_{AB} U^B du dx^A
    + r^2 h_{AB} dx^A dx^B
    \,,
    \label{eqn:BS_stereo_metric}
\end{align}
where~$h^{AB} h_{BC} =
\delta^A_C$,~$\textrm{det}(h_{AB}) = \textrm{det}(q_{AB}) = q$,
with~$q_{AB}$ the metric on the unit sphere. The sphere is
parametrized using the stereographic coordinates~$x^A = (q,p)$
following~\cite{BisGomLeh97a}, though see~\cite{GomLehPap97, Szi00}
for a different but equivalent choice. The metric of the unit sphere
reads
\begin{align*}
q_{AB}dx^A dx^B = \frac{4}{P^2} \left(dq^2 + dp^2 \right)\,,
\end{align*}
where~$P=1+q^2+p^2$. One can introduce a complex basis vector~$q^A$
(dyad)
\begin{align*}
  q^A = \frac{P}{2}\left(1,i\right)
  \,,
\end{align*}
and then the metric of the unit sphere can be written as
\begin{align*}
  q_{AB} = \frac{1}{2} \left(q_A \bar{q}_B + \bar{q}_A q_B \right)
  \,.
\end{align*}

Using the complex dyad, a tensor
field~$F_{A_1 \dots A_n}$ on the sphere can be represented as
\begin{align*}
  F = q^{A_1} \dots q^{A_p} \bar{q}^{A_{p+1}} \dots \bar{q}^{A_n} F_{A_1 \dots A_n}
  \,,
\end{align*}
which obeys the relation~$F \rightarrow e^{i s \psi}F$, with spin
weight~$s=2p-n$. The eth operators for this quantity are defined as
\begin{align*}
  \eth F \equiv q^A \nabla_A F
  &= q^A \p_A F + \Gamma s F
    \,,
  \\
  \bar{\eth} F \equiv \bar{q}^A \nabla_A F
  &= \bar{q}^A \p_A F - \bar{\Gamma} s F
    \,,
\end{align*}
with spin~$s \pm 1$, respectively, and~$\nabla_A$ the covariant
derivative associated with~$q_{AB}$,
i.e.,~$\Gamma = -\frac{1}{2}q^a \bar{q}^b \nabla_a q_b$. In the chosen
stereographic coordinates the above reads
\begin{align*}
  \eth F
  &=
    \frac{P}{2}
    \p_q F
    +
    i \frac{P}{2}
    \p_{p} F
    + \left(q + i p \right) s F
    \,,
  \\
  \bar{\eth} F
  &=
    \frac{P}{2}
    \p_q F
    -
    i \frac{P}{2}
    \p_p F
    - \left(q - i p \right) s F
    \,.
\end{align*}

It is convenient to introduce the following complex spin-weighted
quantities:
\begin{align*}
  J \equiv \frac{h_{AB} q^A q^B}{2} \,,
  \quad
  K \equiv \frac{h_{AB} q^A \bar{q}^B}{2}
  \, ,
  \quad
  U \equiv U^A q_A
  \,,
\end{align*}
as well as the real variable
\begin{align*}
  W \equiv \frac{V-r}{r^2}
  \,.
\end{align*}
Because of the determinant
condition~$\textrm{det}(h_{AB}) = \textrm{det}(q_{AB})$ the
quantities~$K$ and $J$ are related via~$1 = K^2 - J \bar{J}$.~$J$ has
spin-weight two,~$U$ one, and~$K$,~$W$,~$\beta$ zero. The spin weight
of the complex conjugate is equal in magnitude and opposite in
sign. To eliminate second radial derivatives of~$U$ the following
intermediate quantity is introduced:
\begin{align*}
  Q_A \equiv r^2 e^{-2 \beta} h_{AB} U^B_{,r}
  \,.
\end{align*}
Using these variables, the implemented vacuum EFE consist of the
hypersurface equations
\begin{subequations}
\begin{align}
  \beta_{,r}
  & = N_\beta
  \,,
  \label{eqn:hypersurf_beta}
  \\
  \left( r^2 Q \right)_{,r}
  & =
    -r^2 \left( \bar{\eth} J + \eth K \right)_{,r}
    \nonumber
  \\
  & \qquad \qquad 
  +2 r^4 \eth \left( r^{-2} \beta \right)_{,r}
  + N_Q \,,
    \label{eqn:hypersurf_Q}
  \\
  U_{,r}
  & =
  r^{-2} e^{2 \beta} Q 
  + N_U \,,
    \label{eqn:hypersurf_U}
  \\
  W_{,r}
  & =
    \frac{1}{2} e^{2\beta} \mathcal{R}
    -1 - e^\beta \eth \bar{\eth} e^\beta
    \nonumber \\
  & \qquad
    + \frac{1}{4} r^{-2} \left[
    r^4 \left(\eth \bar{U} + \bar{\eth} U \right)
    \right]_{,r}
  + N_W \,,
    \label{eqn:hypersurf_W}
\end{align}
\label{eqn:hypersurf_sys}%
\end{subequations}
where~$Q\equiv Q_A q^A$ and
\begin{align*}
  \mathcal{R}
  &= 2 K - \eth \bar{\eth} K
    + \frac{1}{2} \left(\bar{\eth}^2 J + \eth^2 \bar{J} \right)
    \nonumber
  \\
  & \qquad \qquad \qquad
  + \frac{1}{4K} \left(
  \bar{\eth}\bar{J} \eth J - \bar{\eth}J \eth \bar{J}
  \right)
  \,,
\end{align*}
the curvature scalar for surfaces of constant~$u$ and~$r$. The
evolution equation of the system is
\begin{align}
  &
    2 \left(r J\right)_{,ur} -
  \left[ \frac{r+W}{r} \left( rJ \right)_{,r} \right]_{,r}
    =
    -r^{-1} \left( r^2 \eth U \right)_{,r}
    \nonumber \\
  & \qquad \qquad \quad
     + 2r^{-1} e^\beta \eth^2 e^\beta
    - J \left( r^{-1} W \right)_{,r}
    + N_J \,.
    \label{eqn:evol_J}
\end{align}
The complete form of~$N_\beta\,, N_Q\,, N_U\,, N_J$ in terms of the
eth formalism can be found in~\cite{ReiBisPol12}. The
system~\eqref{eqn:hypersurf_sys} and~\eqref{eqn:evol_J} corresponds to
the main equations~\eqref{eqn:main_BS_sys} in the Bondi-Sachs proper
gauge~\eqref{eqn:BS_stereo_metric}. A pure gauge analysis of this
system was presented in Sec.~\ref{Subsection:BS_proper_gauge}. For
comparison purposes we employ also the following {\it artificial}
symmetric hyperbolic system
\begin{subequations}
\begin{align}
  \beta_{,r}
  & = N_\beta
  \,,
  \\
  \left( r^2 Q \right)_{,r}
  & =
    0 \,,
  \\
  U_{,r}
  & =
  r^{-2} e^{2 \beta} Q 
  + N_U \,,
  \\
  W_{,r}
  & =
    0 \,,
  \\
    2 \left(r J\right)_{,ur}
    &=
      \left[ \frac{r+W}{r} \left( rJ \right)_{,r} \right]_{,r}
        \,.
\end{align}
\label{eqn:SH_artificial_sys}%
\end{subequations}

Equations~\eqref{eqn:hypersurf_W} and~\eqref{eqn:evol_J} involve the
conjugate variables~$\bar{U}$ and~$\bar{J}$, for which the
system~\eqref{eqn:hypersurf_sys} and~\eqref{eqn:evol_J} does not
explicitly possess evolution equations. For the hyperbolicity analysis
provided in the ancillary files we need to complete the system in the
sense of having one equation for each variable. We obtain the
equations for~$\bar{U}$,~$\bar{Q}$ and~$\bar{J}$ by taking the complex
conjugate of~\eqref{eqn:hypersurf_Q},~\eqref{eqn:hypersurf_U}
and~\eqref{eqn:evol_J}, respectively. The state vector of the
linearized about Minkowski and first order reduced system is
\begin{align*}
  \mathbf{u}
  &= \left(
  \beta\,, \beta_q \,, \beta_p \,,
  Q\,, \bar{Q}\,,
  U\,, U_q\,, U_p\,,
  \bar{U}\,, \bar{U}_q\,, \bar{U}_p\,,
   \right. \nonumber \\
  & \qquad \left. W\,,
  J\,, J_r\,, J_q\,, J_p\,,
  \bar{J}\,, \bar{J}_r\,, \bar{J}_q\,, \bar{J}_p\,,
  \right)^T\,,
\end{align*}
where
\begin{align*}
  & \beta_q \equiv \p_q \beta\,,
    \quad
    \beta_p \equiv \p_p \beta\,,
    \quad
    U_q \equiv \p_q U \,,
    \quad
    U_p \equiv \p_p U \,,
    \nonumber
  \\
  & J_q \equiv \p_q J \,,
    \quad
    J_p \equiv \p_p J \,,
    \quad
    J_r \equiv \p_r J
    \,,
\end{align*}
and the complex conjugates are defined in the obvious way. In the ADM
coordinates~$(t,\rho,p,q)$ with
\begin{align*}
  u = t-\rho \, , &\qquad r=\rho \,,
\end{align*}
the system can be written in the form
\begin{align*}
  \p_t \mathbf{u} 
  + \mathbf{B}^\rho \, \p_\rho  \mathbf{u}
  + \mathbf{B}^q \, \p_q  \mathbf{u} \,
  + \mathbf{B}^p \, \p_p  \mathbf{u}
  + \mathbf{S} = 0
  \, .
\end{align*}
Just as the systems analyzed in~\cite{GiaHilZil20} it is only WH due
to the nondiagonalizability of the principal symbol along the angular
directions~$q$ and~$p$. The characteristic variables along the radial
direction with speed~$-1$ are ingoing and consist of
\begin{align*}
  \frac{J}{r} + J_r
  \,,
\end{align*}
and its complex conjugate. The outgoing variables are those with
speed~$1$, namely
\begin{align*}
  &- \frac{J}{r}
    \,,
    \quad
    J_q
    \,,
    \quad
    J_p
    \,,
    \quad \;
    U
    \,,
    \quad
    U_q
    \,,
    \; \; \;
    U_p
    \,,
  \\
  &
    \quad \;
    Q
    \,,
    \quad \;
    W
    \,,
    \quad \; \;
    \beta
    \,,
    \quad \,
    \beta_q
    \,,
    \; \; \; \,
    \beta_p
    \,,
\end{align*}
and their appropriate complex conjugates.

In analogy to the characteristic toy models of~\cite{GiaHilZil20}, we
perform norm convergence tests where the ingoing variables are
integrated over a null hypersurface and the outgoing ones over a world
tube of constant radius. The code works with the compactified radial
coordinate
\begin{align*}
  z = \frac{r}{R_E + r}
  \,,
\end{align*}
where~$R_E$ is a constant that denotes the extraction radius and for
our tests we set it equal to one. If the grid spacing is denoted
as~$h_z,h_q,h_p$ for the coordinates~$z,q,p$, respectively, and the
time step as~$h_u$, then the discretized version of the~$L^2$ norm
that we use is
\begin{equation}
  \begin{aligned}
    &
    ||\mathbf{u}_h|| =
    \left\{
      \sum_{z,q,p} \,
    \left[
      \left(\frac{J}{r} + J_r\right) \left(\frac{\bar{J}}{r} + \bar{J}_r\right) 
    \right]
    \, h_z \, h_q \, h_p 
  \right\}^{1/2}
  \\
    & \quad  
    + \textrm{max}_z
    \left\{
      \sum_{u,q,p} \,
    \left(
      \beta^2  + \beta_q^2 + \beta_p^2 + W^2+
    Q \bar{Q} +
    U \bar{U}
  \right. \right.
    \\
    &
    \quad
    \left. \left.
      + U_q \bar{U_q} + U_p \bar{U_p}
      + \frac{J\bar{J}}{r^2} + J_q \bar{J}_q + J_p \bar{J}_p
    \right)
    h_u \, h_q \, h_p
  \right\}^{1/2}
  \,,
  \end{aligned}
  \label{eqn:discrete_L2_car_vars}
\end{equation}
where the functions in the sums are to be understood as grid
functions. All the outgoing variables of the artificial SH
system~\eqref{eqn:SH_artificial_sys} satisfy advection equations
toward future null infinity. We further introduce
\begin{align*}
  U_q\, \quad U_p \,, \quad
  \beta_q\,, \quad \beta_p\,,
\end{align*}
as well as the appropriate complex conjugates as independent
variables, even though it is not necessary, in order to include in the
norm terms with angular derivatives. These variables are also
outgoing, and their equations of motion are obtained by acting with
the appropriate derivatives to those of~$U$,~$\bar{U}$,
and~$\beta$. Consequently, the appropriate~$L^2$ norm for this system
is~\eqref{eqn:discrete_L2_car_vars} without the terms~$J_q \bar{J}_q$
and~$J_p \bar{J}_p$.

\subsection{Convergence tests}

\begin{figure*}[!t]
  \includegraphics[width=1\textwidth]{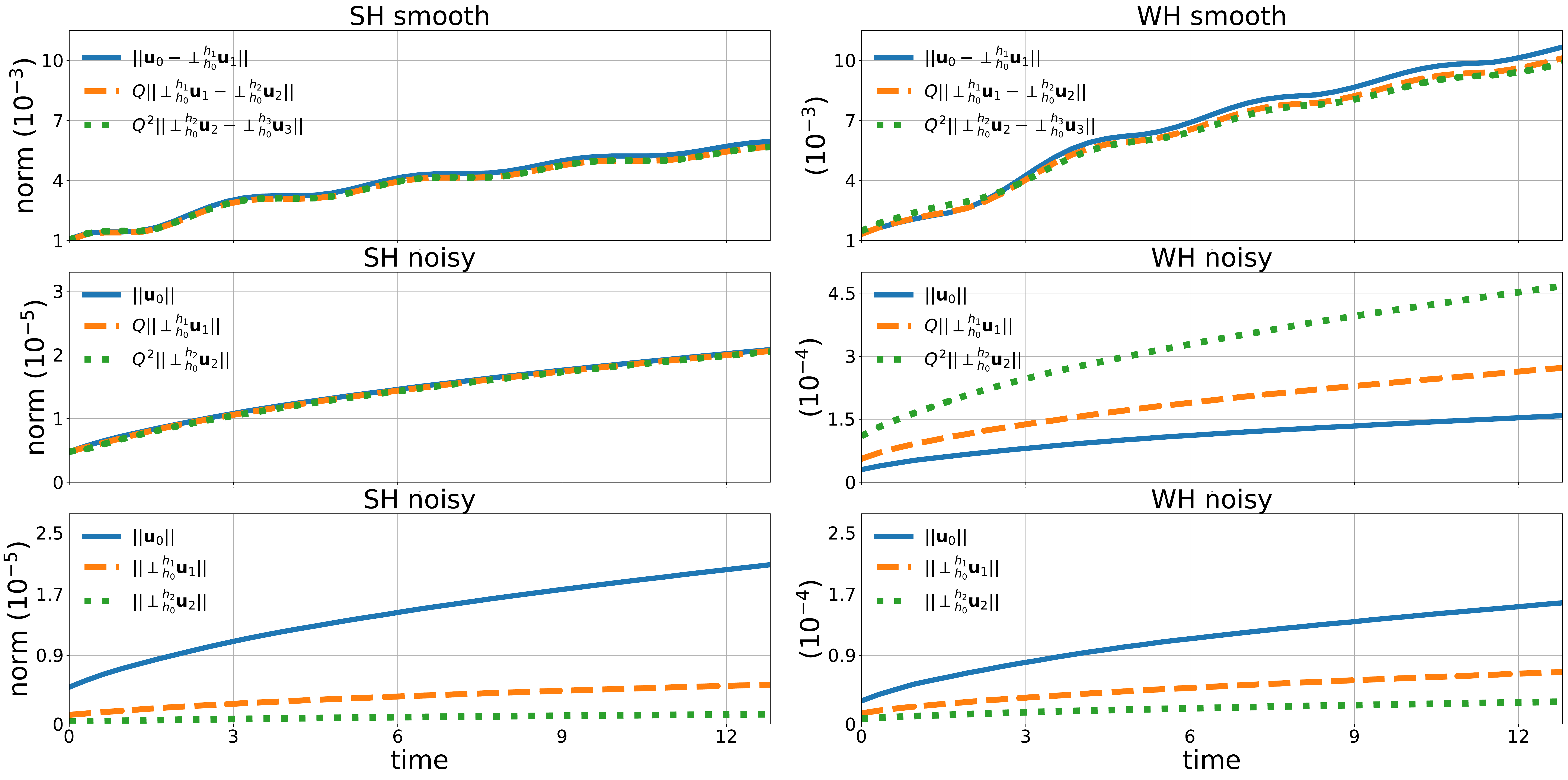}
  \caption{Self (above) and exact (below) convergence tests for the
    artificial SH system and the full Bondi-Sachs system that is
    WH. In the top and middle rows the rescaled norms are shown, with
    rescaling factor~$Q=4$. The overlap of the rescaled norms is
    understood as convergence and the lack of overlap as
    nonconvergence. The tests are performed in the
    norm~\eqref{eqn:discrete_L2_car_vars} for the WH and the
    norm~\eqref{eqn:discrete_L2_car_vars} without
    the~$J_q\bar{J}_p + J_p\bar{J}_p$ term for the SH system. The
    self-convergence tests with smooth data are passed by both
    systems. The exact convergence tests with noisy data are passed
    only by the SH system. In the middle right subfigure we see the
    failure of convergence of the full Bondi-Sachs system, as expected
    by theory. In the bottom row the original norms without rescaling
    are shown. This illustrates that even though the numerical error
    converges to zero with increasing resolution also for the WH case,
    the rate at which this happens is not the expected one, and this
    is understood as loss of convergence.}
  \label{Fig:rescaled_norms_all}
\end{figure*}

In the convergence tests we solve the same PDE problem with increasing
resolution and we monitor the behavior of the numerical error. The
numerical domain is
\begin{align*}
  u \in [0,12.8]
  \,, \quad
  z \in [0.45,1]
  \,, \quad
  p,q \in [-2,2]
  \,,
\end{align*}
where~$u$ denotes time,~$z$ is the compactified radial coordinate,
and~$p,q$ the angular coordinates. The two-sphere is covered by
overlapping north and south patches. In the parameter files included
in the Supplemental Material~\cite{GiaBisHil21_public} the
variables~$y,x$ correspond to the~$p,q$ angular coordinates. These
variables refer to the \texttt{Einstein Toolkit} thorn
\texttt{CartGrid3D} and their domain size is different. The grid they
provide corresponds to the grid for~$p,q$. As described
in~\cite{BabBisSzi09}, the~$p,q$ grid points are
\begin{align*}
  p_i
  &= -1 + \Delta (i - O - 1)
    \,,
  \\
  q_j
  &= -1 + \Delta (j - O - 1)
    \,,
\end{align*}
where~$O$ denotes the number of overlapping points beyond the
equator. The range of the indices is
\begin{align*}
  1 \leq i,j \leq M + 1 + 2O
  \,,
\end{align*}
where~$M^2$ is the total number of $p,q$ grid points inside the
equator and~$\Delta = 2/M$ is the grid spacing. The physical part of
the stereographic domain consists of the grid points for which
\begin{align*}
  p^2 + q^2 \leq 1
  \,,
\end{align*}
and these are the only points considered in our tests. We
label the different resolutions as~$h_0,\, h_1,\, h_2,\, h_3$ with
\begin{align*}
  h_0: \quad
  N_z,N_p,N_q
  &= 33 \,,
    \quad \; \,
    h_u = 0.04
    \,,\\
  h_1: \quad
  N_z,N_p,N_q
  &= 65 \,,
    \quad \; \,
    h_u = 0.02
    \,,\\
  h_2: \quad
  N_z,N_p,N_q
  &= 129 \,,
    \quad
    h_u = 0.01
    \,,\\
  h_3: \quad
  N_z,N_p,N_q
  &= 257 \,,
    \quad
    h_u = 0.005
    \,,
\end{align*}
and~$N_z,N_p,N_q$ the number of points in the~$z,p,q$ numerical grids.
$N_p,N_q$ refer to the total number of grid points (overlapping and
nonoverlapping regions together). By construction the grid points and
time steps of~$h_0$ are common for all resolutions.

We perform convergence tests using both smooth and noisy given
data. The former are based upon the linearized gravitational wave
solutions derived in~\cite{Bis05} and adapted to the notation used
here in~\cite{ReiBisLai06, BabBisSzi09}, namely
\begin{align*}
  J
  &= \sqrt{(l-1) l (l+1) (l+2)} {}_2 R_{lm} \Re (J_l(r) e^{i \nu u})
  \,,
  \\
  U
  & = \sqrt{l (l+1)} {}_1 R_{lm} \Re (U_l(r) e^{i \nu u})
    \,,
      \\
  \beta
  &= R_{lm} \Re ( \beta_l e^{i \nu u})
    \,,
      \\
  W_c
  & = R_{lm} \Re ( W_{cl}(r) e^{i \nu u})
    \,,
\end{align*}
where~$W_c$ gives the perturbation to~$V$ and for~$l=2$
\begin{align*}
  \beta_2
  & = \beta_0
    \,
  \\
  J_2(r)
  & = \frac{24 \beta_0 + 3 i \nu C_1 - i \nu^3 C_2}{36}
    + \frac{C_1}{4r}
    - \frac{C_2}{12 r^3}
    \,,
  \\
  U_2(r)
  &= \frac{-24 i \nu \beta_0 + 3 \nu^2 C_1 - \nu^4 C_2}{36}
    + \frac{2 \beta_0}{r}
    + \frac{C_1}{2 r^2}
  \\
  & \qquad
    + \frac{i \nu C_2}{3 r^3}
    + \frac{C_2}{4 r^4}
    \,,
  \\
  W_{c2}(r)
  & =
    \frac{24 i \nu \beta_0 - 3 \nu^2 C_1 + \nu^4 C_2}{6}
  - \frac{\nu^2 C_2}{r^2}
    \\
  & \qquad
    + \frac{3 i \nu C_1 - 6 \beta_0 - i \nu^3 C_2}{3 r}
    + \frac{i \nu C_2}{r^3}
    + \frac{C_2}{2 r^4}
    \,.
\end{align*}
We fix the parameters of these solutions to
\begin{align*}
    \nu &= 1
    \,, \quad
    l = 2
    \,, \quad
    m = 0 \,,
    \\
    C_1 &= 3 \cdot 10^{-3}
    \,, \quad
    C_2 = 10^{-3}
    \,, \quad
    \beta_0 = i \cdot 10^{-3}
    \,.
\end{align*}
The constant~$\nu$ controls the frequency of the solution,~$l,m$ refer
to the spin-weighted spherical harmonics and~$C_1, C_2, \beta_0$ are
integration constants.

For the noisy tests we set all the initial and boundary data to their
Minkowski values, perturbed with random noise of amplitude~$A$ with
\begin{align*}
  A(h_0)
  &= 4096 \cdot 10^{-10}
  \,, \quad
  A(h_1)
  = 512 \cdot 10^{-10}
  \,, \quad
  \\
  A(h_2)
  &= 64 \cdot 10^{-10}
  \,,
\end{align*}
on all the given data. The scaling of the amplitude by a factor of 8
every time we double resolution is due to the first order derivatives
in the norm~\eqref{eqn:discrete_L2_car_vars}, as explained in Sec. IV
of~\cite{GiaHilZil20}. The amplitude of the noise is low enough for
the nonlinear terms to be negligible with the precision at which we
work. The complete parameter files used in the simulations can be
found in the ancillary files. We call self-convergence the tests in
which we obtain an error estimate by taking the difference between two
numerical solutions. This is useful when an exact solution is not
known, as, for instance, for the artificial SH
system~\eqref{eqn:SH_artificial_sys} when smooth data are
given. Hence, we perform self-convergence tests in the smooth setup
for both WH and SH systems. On the contrary, the noisy tests consist
of random noise on top of vanishing given data for both systems and
zero is a solution for both cases. So, for this case we perform exact
convergence tests, i.e., the error estimate is provided by a
comparison between the numerical and the exact solutions. We use the
operator~$\perp_{h_0}^{h_i}$ to denote that we consider only the
common grid points of the resolution~$h_i$ with the coarse
resolution~$h_0$, as well as the common time steps. For the
self-convergence tests we monitor
\begin{align*}
  &
    ||\mathbf{u}_{h_0} - \perp_{h_0}^{h_1} \mathbf{u}_{h_1}||
  \,, \qquad
  || \perp_{h_0}^{h_1}\mathbf{u}_{h_1} - \perp_{h_0}^{h_2} \mathbf{u}_{h_2}||
  \,,
  \\
  &
    || \perp_{h_0}^{h_2}\mathbf{u}_{h_2} - \perp_{h_0}^{h_3} \mathbf{u}_{h_3}||
    \,,
\end{align*}
and for the exact convergence
\begin{align*}
  ||\mathbf{u}_{h_0}||
  \,, \quad
  || \perp_{h_0}^{h_1} \mathbf{u}_{h_1}||
  \,, \quad
  || \perp_{h_0}^{h_2} \mathbf{u}_{h_2}||
  \,.
\end{align*}
The code uses finite difference operators that are second order
accurate. This, combined with the doubling of grid points every time
we increase resolution provides a convergence
factor~$Q=4$~\cite{GiaHilZil20}. 

\begin{figure*}[!t]
  \includegraphics[width=0.985\textwidth]{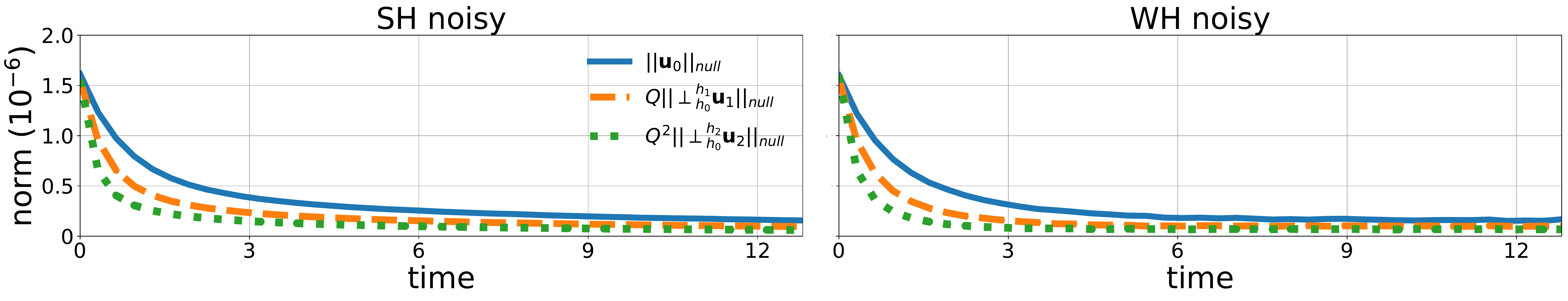}
  \caption{Exact convergence test with noisy data for both PDE
    systems, using only the null part of the
    norm~\eqref{eqn:discrete_L2_car_vars}. The WH system does not
    manifest a clear loss of convergence. Similar to~\cite{HanSzi14}
    there is no evidence of exponential growth.}
  \label{Fig:null_noisy_rescaled_norms}
\end{figure*}

In Fig.~\ref{Fig:rescaled_norms_all} the rescaled norms for both
smooth and noisy tests, for the artificial
SH~\eqref{eqn:SH_artificial_sys} and the full Bondi-Sachs
system~\eqref{eqn:hypersurf_sys} and~\eqref{eqn:evol_J} that is WH are
illustrated. The overlap of the rescaled norms indicates good second
order convergence, whereas the lack of overlap suggests
nonconvergence.  For smooth given data both the SH and the WH systems
exhibit good decent order convergence. However, for noisy given data
only the SH has the appropriate convergence. This feature is expected,
as noisy given data are important to demonstrate WH in numerical
experiments~\cite{CalHinHus05, GiaHilZil20}. These results are
compatible with earlier tests with random noise that demonstrated the
lack of exponential growth in the solution~\cite{HanSzi14}. In
Fig.~\ref{Fig:null_noisy_rescaled_norms} the sum only over the null
hypersurface from~\eqref{eqn:discrete_L2_car_vars} is shown, that is
similar to earlier tests. The loss of convergence in the WH system is
less severe than for the full norm~\eqref{eqn:discrete_L2_car_vars},
and there is no sign of exponential growth in the solution. This fact
alone may be evidence for numerical stability in the colloquial sense
that the code does not crash but, as we demonstrate in
Fig.~\ref{Fig:rescaled_norms_all}, is not enough evidence for
convergence. It becomes apparent then that the choice of norm in which
the convergence tests are performed is crucial. A norm that is
compatible with the PDE system under consideration should be used.

\section{Conclusions}
\label{Section:Conclusions}

Characteristic formulations of GR are used in a number of cases such
as gravitational waveform modeling, critical collapse and applications
to holography. These formulations are most commonly built upon
Bondi-like gauges. In~\cite{GiaHilZil20} the EFEs were shown to be
only weakly hyperbolic in second order metric form in two popular
Bondi-like gauges. Computational experiments were performed on toy
models to examine the consequences of this fact, with the conclusion
that numerical convergence in the simplest desired norms does not
occur. Building on~\cite{GiaHilZil20}, in this paper we showed that
this weak hyperbolicity is caused by the gauge condition~$g^{uA}=0$
common to all Bondi-like gauges. Subsequently we performed numerical
experiments performed in full GR, and found that the conclusions
of~\cite{GiaHilZil20} indeed carry over; ill-posedness of the
continuum PDE (in the natural equivalent of~$L^2$) for the
characteristic problem serves as an obstruction to convergence of the
numerics (in a discrete approximation to the same norm).

To show that weak hyperbolicity was a pure gauge effect we had to jump
through a number of technical hoops. We mapped the characteristic free
evolution system to an ADM setup so that the results
of~\cite{KhoNov02,HilRic13} could easily be used. This allowed us to
distinguish among the gauge, constraint, and physical degrees in the
linear, constant coefficient approximation. Crucially it is known that
weakly hyperbolic pure gauges give rise to weakly hyperbolic
formulations. We were able to show the former in a number of cases.
Specifically, we have studied three Bondi-like setups: the affine
null, the Bondi-Sachs proper and the double null gauges. All three
have the same degenerate structure rendering the pure gauge subsystem
weakly hyperbolic. We have thus argued that when the EFE are written
in a Bondi-like gauge with at most second derivatives of the metric
and there are nontrivial dynamics in at least two spatial directions,
then, due to the weak hyperbolicity of the pure gauge subsystem, the
resulting PDE system is only WH.

The implication of weak hyperbolicity is that the CIVP and CIBVP of GR
are ill-posed in the natural equivalent of~$L^2$ on these geometric
setups. Therefore we carried out convergence tests in a discretized
version of such a norm. The specific form of the norm is inspired by
the characteristic toy models of~\cite{GiaHilZil20}. We performed the
tests on the Bondi-Sachs gauge system~\eqref{eqn:hypersurf_sys}
and~\eqref{eqn:evol_J} implemented in the~\texttt{PITTNull} thorn of
the~\texttt{Einstein Toolkit}, as well as on the artificial strongly
hyperbolic system~\eqref{eqn:SH_artificial_sys}. The norm used is
compatible with the strongly hyperbolic model in the characteristic
domain. The tests are performed with smooth and with noisy given
data. For smooth data both the strongly and weakly hyperbolic systems
model exhibit good convergence. But with noisy data only the strongly
hyperbolic model retains this behavior. These findings are compatible
with previous results~\cite{CalHinHus05, CaoHil11, GiaHilZil20},
namely that noisy given data are essential to reveal weak
hyperbolicity in numerical experiments. We have furthermore seen that
even with noisy data one might overlook this behavior if tests are
performed in a norm that is not suited to the particular problem.

Given all of the above, the obvious approach to circumvent weak
hyperbolicity is to adopt a different gauge. For applications in CCM
this may be necessary, since it is otherwise not at all clear how a
well-posedness result for the composite PDE problem could be
obtained. Yet, as discussed in the Introduction, concerning purely
characteristic evolution, symmetric hyperbolic formulations of GR
employing Bondi-like gauges are
known~\cite{Rac13,CabChrTag14,HilValZha19,Rip21}. At first sight this
seems to contradict the claim that any formulation of GR inherits the
pure gauge principal symbol within its own. But these formulations all
promote the curvature to be an evolved variable, so the results
of~\cite{HilRic13} do not apply. As we have seen in
Sec.~\ref{Section:aff_null_gauge}, taking an outgoing null derivative
of the affine null pure gauge subsystem, we obtain a strongly
hyperbolic PDE. It is thus tempting to revisit the model
of~\cite{HilRic13} to investigate the conjecture that formulations of
GR with evolved curvature can be built that inherit specific
derivatives of the pure gauge subsystem. A deeper understanding of the
relation between the latter and the Bondi-like formulations analyzed
in this paper could suggest norms in which they are actually
well-posed. Obtaining such a proof would help validate error estimates
for numerical solutions so relevant for applications in gravitational
wave astronomy. Work in this direction is ongoing and will be reported
on elsewhere.

\acknowledgments

We are grateful to Fernando Abalos, Carsten Gundlach and Justin Ripley
for helpful discussions. Supplemental Material and our data can be
found in the ancillary files and in~\cite{GiaBisHil21_public}. The
work was partially supported by the FCT (Portugal) IF
Program~IF/00577/2015, IF/00729/2015, PTDC/MAT-APL/30043/2017 and
Project~No.~UIDB/00099/2020. T.G.  acknowledges financial support
provided by FCT/Portugal Grant No. PD/BD/135425/2017 in the framework
of the Doctoral Programme IDPASC-Portugal. N.T.B. was supported by the
National Research Foundation, South Africa, under Grant No. 118519.
The authors acknowledge networking support by the GWverse COST Action
CA16104, ``Black holes, gravitational waves and fundamental physics''.

\bibliography{refs}

\end{document}